\documentclass[times,review,preprint,authoryear]{elsarticle}

\usepackage{jasr}
\usepackage{framed,multirow}

\usepackage{amsmath,amssymb}
\usepackage{latexsym}
\usepackage{siunitx}
\usepackage{gensymb,wasysym}
\usepackage{caption,subcaption}

\usepackage{url}
\usepackage{xcolor}
\definecolor{newcolor}{rgb}{.8,.349,.1}

\usepackage[noprefix]{nomencl}
\makenomenclature

\newcommand{\phoenix}{\emph{Phoenix} }

\usepackage{tabularx,longtable}

\usepackage[citebordercolor=white]{hyperref}
\usepackage[capitalise]{cleveref}
\crefrangelabelformat{equation}{(#3#1#4-#5#2#6)}
\crefformat{appendix}{#2#1#3}

\journal{Advances in Space Research}

\begin{document}

\verso{Mirko Trisolini \textit{etal}}

\begin{frontmatter}

\title{Re-entry prediction and demisability analysis for the atmospheric disposal of geosynchronous satellites}

\author[1]{Mirko \snm{Trisolini}\corref{cor1}}
\cortext[cor1]{Corresponding author}
\ead{mirko.trisolini@polimi.it}
\author[1]{Camilla \snm{Colombo}}
\ead{camilla.colombo@polimi.it}

\address[1]{Politecnico di Milano, Department of Aerospace Science and Technology, Via La Masa 34, 20156, Milano, Italy}


\begin{abstract}
	The paper presents a re-entry analysis of Geosynchronous Orbit (GSO) satellites on disposal trajectories that enhance the effects of the Earth oblateness and lunisolar perturbations. These types of trajectories can lead to a natural re-entry of the spacecraft within 20 years. An analysis was performed to characterise the entry conditions for these satellites and the risk they can pose for people on the ground if disposal via re-entry is used. The paper first proposes a methodology to interface the long-term propagation used to study the evolution of the disposal trajectories and the destructive re-entry simulations used to assess the spacecraft casualty risk. This is achieved by revisiting the concept of overshoot boundary. The paper also presents the demisability and casualty risk analysis for a representative spacecraft configuration, showing that the casualty risk is greater than the 10$^{-4}$ threshold and that further actions should be taken to improve the compliance of these satellites in case disposal via re-entry is used.
\end{abstract}


\end{frontmatter}


\section{Introduction}
\label{sec:intro}

In the last decade, the number of artificial satellites launched by private and governmental agencies increased exponentially. The growth of space activities results in systematic congestion of specific orbital regions about the Earth. Nowadays, the Geosynchronous Orbit (GSO) region and the Low Earth Orbit (LEO) region are the two most populated. The importance of the LEO region resides in the cheaper accessibility for satellite services, such as Earth observations, remote sensing, and telecommunication services. On the other hand, satellites in the GSO region provide a fundamental contribution to global communication, television broadcasting and weather forecasting services.
A significant consequence of the increase in space activity is the growth of inactive satellites still in orbit, which become space debris. The Inter-Agency Space debris Coordination Committee (IADC) defines the recommended guidelines for the mitigations and decommissioning of defunct spacecraft \citep{IADC2011}. While the disposal of any object in LEO is classically achieved by natural re-entry in a short time due to atmospheric drag, satellites in GSO, Medium Earth Orbit (MEO) or {Highly Elliptical Orbit (HEO) should implement different strategies. The guideline for ESA-operated GSO missions is to increase the altitude of perigee by 235 km (considering a correction factor for the solar radiation pressure perturbations) and to circularise the graveyard orbit \citep{ESASpaceDebrisMitigation2015}. Different approaches have been proposed in the literature to reduce collision probability in these regions. Lidov and Kozai \citep{lidov1963evolution,kozai1962secular} applied Hamiltonian theory to assess the secular variations caused by the third body effect. The idea of exploiting the lunisolar gravitational perturbation for for satellite disposal was suggested for satellites in MEO region \citep{alessi2016numerical,skoulidou2017cartographic,armellin2018optimal} and HEO region  \citep{jenkin2008lifetime,Colombo2019surf,colombo2019long}. The end-of-life strategy of the INTEGRAL mission also exploited the long-term dynamics combining the effect of natural perturbations and an optimal impulsive manoeuvre to design a safe re-entry trajectory from HEO \citep{colombo2014end,armellin2015end,merz2015orbit}. 

Gkolias et al. \citep{Gkolias2019GEO}  presented a similar promising strategy for satellites in the geosynchronous region. The idea is to exploit lunisolar perturbations to implement long-term disposal trajectories for highly inclined GSO missions, which have a distinct dynamical behaviour under the lunisolar effect \citep{delhaise1993luni,wytrzyszczak2007regular,zhang2017long}. \cite{Gkolias2019GEO} demonstrated the feasibility of post-mission disposal strategies for the GSO region. These strategies are related to particular orbital geometries about the Moon and Sun orbital plane, which exploit the Lidov-Kozai dynamics resonances to achieve atmospheric re-entry within 20 years for highly-inclined orbits. Alongside the lunisolar effect, the semi-analytical dynamical model included the main perturbing terms for the GSO region: the geopotential perturbations, including the zonal and resonant tesseral harmonics, the solar radiation pressure and the Earth's general precession. The single-averaged theory for Earth's satellite was used, implemented in the PlanODyn suite, derived by \cite{colombo2016planetary}

\bigbreak
The possibility to exploit the resonances with the lunisolar perturbations has opened new options for the disposal of GSO spacecraft that are based on a mechanism of eccentricity build-up that eventually leads to the atmospheric re-entry of the spacecraft. However, when a satellite re-enters the atmosphere, it must comply with the casualty risk threshold of $10^{-4}$ \citep{OConnor2008_NASA_handbook,johnson2010_nasa_debris,ESASpaceDebrisMitigation2015}. GSO spacecraft are, on average, massive (average launch mass of 4100 \si{\kilo\gram})\citep{UnionofConcernedScientists2017} and usually carry several tanks, which are components known to survive the re-entry process. Therefore, if disposal via re-entry is considered for GSO spacecraft, it is necessary to verify their compliance by performing destructive re-entry analyses. In this paper, the demisability of a reference GSO satellite configuration is performed using the \emph{Phoenix} software, developed at the Politecnico di Milano and University of Southampton \citep{TCH2015_IAC,THC2016_JSSE,THC2017_IAC,TLC18Sensitivity,Trisolini2018_AESCTE,Trisolini2021_JSSE}. It is an object-oriented destructive re-entry code, which performs re-entry simulations and casualty risk estimations. It models the satellite using simplified shapes and computes the aerothermodynamic loads on the spacecraft via engineering correlations.

Alongside the spacecraft configuration, it is also necessary to know the conditions at the entry interface. Given that the re-entry from GSO exploits J$_2$ and lunisolar resonances, disposal orbits tend to be characterised by high-eccentricities at the 120 km boundary; therefore, they lead to entry conditions that are considerably different from the ones typically encountered by LEO satellites. Specifically, a resonant re-entry tends to be characterised by super-circular entry velocities and steeper entry flight-path angles. In addition, given the high energy associated with these disposal orbits, the spacecraft is not guaranteed to re-enter at the first passage in the lower atmosphere, thus performing several passes before entry conditions are met. At each passage in the atmosphere, the orbital energy reduces, leading to a process of eccentricity reduction, which, in some cases, can also lead to circularisation of the disposal orbit. Given the variety of possible entry conditions, it is necessary to identify a suitable integration between the long-term orbit evolution and the destructive re-entry phase. The first phase describes orbit evolution under the influence of the perturbations until a critical condition is reached that defines the re-entry. Traditionally, this condition was defined targeting a pericentre altitude or, equivalently, a limit eccentricity that can ensure the spacecraft re-entry. For example, \cite{Colombo2014integral} selected a 50 km pericentre altitude for the re-entry of INTEGRAL. Using this condition, we can link to the destructive re-entry phase by computing the spacecraft's state at a predefined entry interface (typically 120 km of altitude). With such a procedure, we cannot understand when the entry conditions are met. Therefore, in this paper, we explore the possibility to define the interface between the long-term propagation and the destructive re-entry phases via the concept of the overshoot boundary \citep{Vinh1980}.

The paper is structured as follows. First, a description of the theoretical background of the work and the methodology developed is given in \cref{sec:methodology}. In this section, an outline of the long-term propagation and destructive re-entry methods used for the study is given in \cref{subsec:long-term} and \cref{subsec:reentry}, respectively. \cref{subsec:overshoot} instead describes the concept of overshoot boundary that is used for the interface between the two types of propagation. \cref{sec:reentry_analysis} shows results of the long-term propagation for a selected set of GSO satellites and a summary of their predicted re-entry conditions. \cref{sec:overshoot_application} presents a re-entry prediction analysis performed with the overshoot boundary, comparing its accuracy with the software \emph{Phoenix}. Finally, \cref{sec:demisability}, shows the demisability analysis of a typical GSO spacecraft configuration. Additionally, it discusses the impact of low altitude passages on the aerothermodynamics loads on the spacecraft.


\section{Methodology}
\label{sec:methodology}
The paper proposes a framework for the analysis of the re-entry of spacecraft structures from resonant trajectories by interfacing the long-term propagation phase with the destructive re-entry phase: from the trajectory evolution under the influence of orbit perturbations to the analysis of the fragments and components surviving re-entry and the consequences they can have for people and properties on the ground. This section gives an overview of the methods and software used for the long-term propagation (\cref{subsec:long-term}) and the destructive re-entry (\cref{subsec:reentry}) phases. In addition, it introduces the concept of \emph{overshoot boundary} as a means to predict the re-entry from GSO resonant trajectories (\cref{subsec:overshoot}).


\subsection{Long-term orbit propagation}
\label{subsec:long-term}

The influence of orbital perturbations is essential to analyse the long-term evolution of disposal orbits. Several studies demonstrated how the secular dynamics can be described by mathematical models to exploits resonances for stability and re-entry purposes. Specifically, separating the long-term and the short term variations is convenient for long term propagation. It can be obtained by averaging techniques, which allow a fast propagation of the averaged dynamic of the satellite over the long period. In this context, the influence of an external third body was widely studied in the past for HEO and GSO orbits. The impact of the third body perturbation on highly inclined orbits is known as the Lidov-Kozai effect. \cite{kozai1962secular} described the third body disturbing potential for the secular effect, eliminating the short-periodic term by the Hamiltonian of the system. In the same period, \cite{lidov1963evolution} developed a model to describe the orbit evolution for an Earth’s artificial satellite under the effect of the gravitational perturbation of both the Moon and the Sun. Their studies opened the possibility to use the phase-space to understand the long-term dynamics under external perturbations. The phase space representation allows the identification of equilibrium solutions and transfer maps for manoeuvre design and disposal strategies identification exploiting the secular effect of the perturbation, enabling fuel-efficient disposal manoeuvres \cite{colombo2014end}.

\bigbreak
From the averaging methods, the PlanODyn suite was developed for long-term propagation in perturbed environment \citep{colombo2016planetary}. The suite implements both the single and the double average equations of the dynamics, including the perturbation effects of the Earth zonal Harmonics, the solar radiation pressure, the third body effect of the Sun and the Moon and the secular disturbing effect due to the aerodynamic drag. In this paper, the authors want to provide an interface to the PlanODyn suite for demisability analysis. In fact, the current averaged aerodynamic effect modelled by PlanODyn struggle to predict the precise re-entry condition for highly elliptical orbits. Therefore, we explore the possibility to interface PlanODyn with a destructive re-entry code such as \phoenix that implements a non-average drag model, as described in \cref{subsec:reentry}.

\subsection{GSO dispsal maps}
\label{subsec:disposal-maps}
The theory of long-term orbital evolution was applied by \cite{Gkolias2019GEO} to describe the dynamical environment for GSO, including the geopotential force, the Sun and Moon perturbing forces and the Earth's J$_2$ precession. In his study, the single average formulation is adopted, with the PlanODyn orbital analysis suite \citep{colombo2016planetary}, to characterise the phase space. From the dynamical study, \cite{Gkolias2019GEO} derived a set of disposal maps that describe the dynamical evolution of the orbits over a period of 120 years as a function of the initial inclination, right ascension of the ascending node, $\Omega$, and argument of perigee, $\omega$. Disposal opportunities were identified for inclined GSO orbits, with inclinations ranging from $50^\circ$ to $90^\circ$, for different area-to-mass ratios. It was demonstrated that the natural behaviour of such orbits tends to increase the eccentricity value, up to 0.8-0.9, in a time frame of about 20-30 years, when the satellite could naturally re-enter the atmosphere. The fast re-entry condition was analysed in \cite{Gkolias2019GEO} to study the possibility to re-enter within the 25-year rule that is regularly used for the LEO region \citep{IADC2011}. Another relevant parameter is the time spent by the satellite in the LEO protected region. It was demonstrated that a re-entry from inclined GSO orbits ensures a residence time in the protected region of few days \citep{gkolias2017end,Gkolias2019GEO}.

\bigbreak

The maps obtained by \cite{Gkolias2019GEO} are used as the starting point in this study as they provide an understanding of the dynamical behaviour of GSO spacecraft in resonant trajectories. However, these maps have been obtained without including the effect of drag. Because the critical eccentricity is defined by setting a target pericentre altitude of 50 km, the contribution of drag forces in the neighbourhood of the pericentre can be significant in influencing the characteristics of the re-entry. \cref{fig:reentry_nodrag,fig:reentry_drag} show the comparison between the trajectory evolution of few selected GSO orbits without and with drag, respectively. As expected, the evolution of the trajectory is influenced by drag only in the latest phases of the re-entry. Its most significant effect can be observed in the bottom plot of \cref{fig:reentry_drag} and corresponds to a \emph{circularisation} effect, with a reduction of eccentricity in the last few days of the re-entry.

\begin{figure}[!htb]
	\centering
	\includegraphics[width=3.25in]{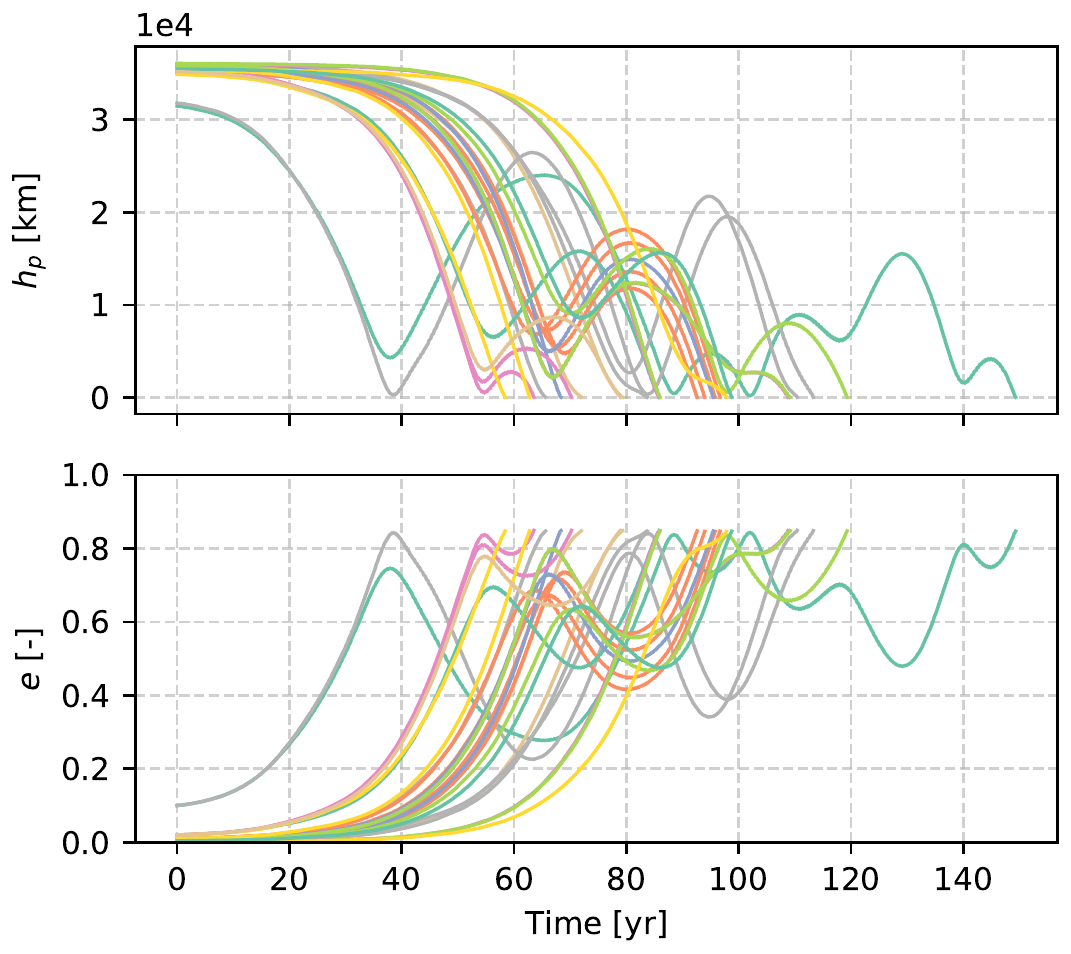}
	\caption{Trajectory evolution of selected GSO orbits without drag. The top plot shows the evolution of the pericentre altitude, while the bottom plot the evolution of the eccentricity. The line colours in the top and bottom plots represent corresponding trajectories.}
	\label{fig:reentry_nodrag}
\end{figure}

\begin{figure}[!htb]
	\centering
	\includegraphics[width=3.25in]{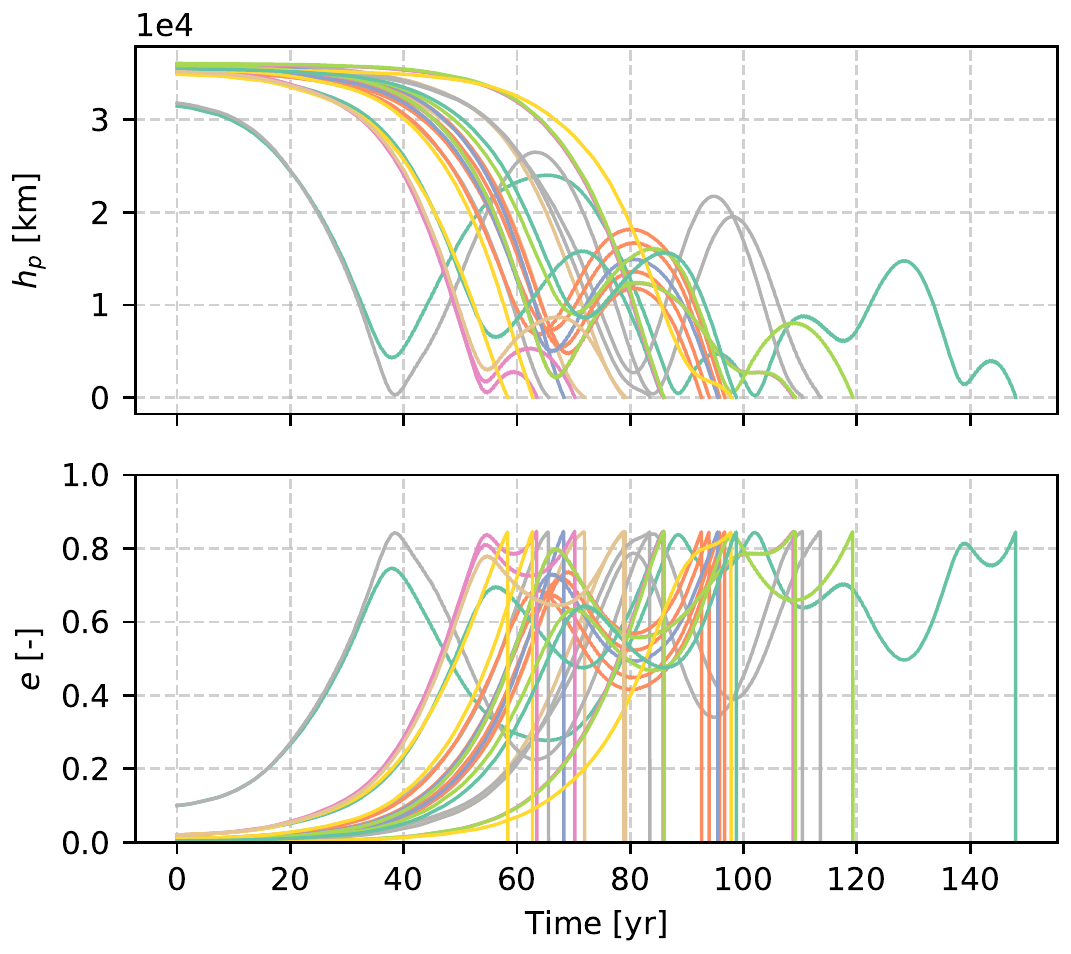}
	\caption{Trajectory evolution of selected GSO orbits with drag. The top plot shows the evolution of the pericentre altitude, while the bottom plot the evolution of the eccentricity. The line colours in the top and bottom plots represent corresponding trajectories.}
	\label{fig:reentry_drag}
\end{figure}

\cref{fig:reentry_drag} shows a case in which all the orbits achieve full circularisation. This result is the effect of imposing in PlanODyn a lower bound for the pericentre altitude of 50 km. Such a low value of the pericentre altitude would cause the spacecraft to experience high thermomechanical loads during its passage through the atmosphere, which would result in its break-up and re-entry. In addition, the assumptions on which PlanODyn is based do not allow for an exhaustive description of the aerothermodynamics to which the spacecraft is subject during this phase, for example, discriminating between free-molecular, transition, and continuum flow. Therefore, to analyse the re-entry from resonant trajectories, such as the geosynchronous ones, it is important to couple the long-term propagation with a destructive re-entry analysis. It is also critical to better understand the switching condition between the two analyses, avoiding the selection of an arbitrary threshold and trying to predict the most suitable switching condition based on the dynamical evolution of the trajectory. 


\subsection{Destructive re-entry}
\label{subsec:reentry}

As mentioned in \cref{sec:intro}, the possibility to dispose of a GSO spacecraft via re-entry also relies on the destructive re-entry and casualty risk analysis of the spacecraft. All re-entering objects must meet the requirement of the casualty risk threshold of $10^{-4}$. Therefore, it is necessary to analyse the spacecraft break-up and demise with destructive re-entry software. In this work, the \emph{Phoenix} software was used. \emph{Phoenix} is an object-oriented code that analyses the re-entry and demise of spacecraft configurations by representing them as a combination of elementary shaped objects (e.g. boxes, spheres, flat plates, disks, and cylinders). The configuration of the spacecraft is defined in a hierarchically, with the \emph{parent} spacecraft representing the main structure, which contains the internal components. The internal components model the main sub-systems such as reaction wheels, tanks, and battery assemblies, and they can contain sub-components to model smaller parts (e.g. battery cells). The internal components are released only after the break-up of the parent spacecraft, which can be triggered by an altitude, a temperature, or a pressure threshold. The entry is modelled via a three-degrees of freedom dynamics, which consider the spacecraft to have a predefined attitude motion (the standard is random tumbling). The gravity model is based on zonal harmonics up to J$_4$, and the Earth's atmosphere uses the 1976 US Standard Atmosphere \citep{1976_std_atm}. The aerodynamics and aerothermodynamics are based on attitude averaged coefficients for the three different flow regimes encountered: free-molecular, transition, and continuum. The ablation of the components uses a lumped mass model, and the material database uses temperature independent materials. In its current version, the \phoenix tool only models the convective heating in the continuum regime; therefore the contribution of the radiative heating is currently not implemented. This last contribution may become relevant for entry speeds above 10 \si{\kilo\meter / \second} \citep{tauber1991stagnation}, as is the case for some of the trajectories analysed in this paper (see \cref{sec:reentry_analysis}). Nonetheless, with the current implementation \emph{Phoenix} provides conservative results, and the casualty risk analysis still provides useful insight for the compliance of GSO satellites with the regulations.

Alongside the re-entry and demise of the components, the on-ground casualty risk can be computed by combining the casualty area, the landing location, and the Gridded Population of the World (GPW) \citep{CIESIN2020} data. Finally, the \emph{Phoenix} suite also supports uncertainty analyses in the initial conditions, break-up conditions, solar panels break-up altitude, and material properties. For further detail on the characteristics of \emph{Phoenix} the reader is referred to \cite{TCH2015_IAC,THC2016_JSSE,THC2017_IAC,TLC18Sensitivity,Trisolini2018_AESCTE,Trisolini2021_JSSE}.


\subsection{Entry interface: the overshoot boundary}
\label{subsec:overshoot}

The present work proposes a methodology to interface the long-term propagation described in \cref{subsec:long-term} with the destructive re-entry analysis outlined in \cref{subsec:reentry}. We propose to use the concept of overshoot boundary, which is related to the definition of the entry corridor of crewed vehicles and has been introduced in the late fifties by Chapman \citep{Chapman1959}. The entry corridor is defined by both an overshoot and an undershoot boundary. In fact, crewed vehicles are required to achieve capture in one pass (remaining below the overshoot boundary) and to avoid high aerothermal loads (remaining above the undershoot boundary). In the case of GSO re-entries, we are only interested in the overshoot boundary so that the spacecraft is captured. The high loads encountered passing below the undershoot boundary are not of interest because the spacecraft is not manned and does not have such requirements. In its simplest definition, the overshoot boundary identifies the maximum pericentre radius that will still allow the spacecraft to be captured in one pass by the planet. The pericentre radius is the one of the approach conic obtained as if the atmosphere would not exist and thus would not influence the motion of the spacecraft.

To compute the overshoot boundary we follow the procedure described by Vinh et al. \citep{Vinh1980,Hicks2009}, where it is assumed that the entry occurs when the deceleration due to the aerodynamic forces reaches a specific fraction, $f$, of the gravitational acceleration. This assumption translates into a condition for which the minimum acceleration along the overshoot boundary is equal to $f = a_{\rm decel} / g_0$. The value of $f$ is somewhat arbitrary; therefore, we decided to follow Vinh's indication and use a value of $f = 0.05$ that is an aerodynamic deceleration whose magnitude is 5\% the gravitational acceleration at sea level. To find the overshoot boundary, Vinh makes use of two adimensional variables, the \emph{modified Chapman's variables}, which are defined as:

\begin{align}
	Z &= \rho(r) \cdot \frac{C_D S}{2 m} \cdot \sqrt{\frac{r}{\beta}}  \label{eq:chapman_var_z} \\
	u &= \frac{v^2 \cdot \cos{\gamma}^2}{g(r) \cdot r},  \label{eq:chapman_var_u}
\end{align}

where $\rho$ is the density of the atmosphere at the radial position $r$, $C_D$ is the drag coefficient, $S$ is the cross-sectional area, $m$ is the mass of the spacecraft, $\beta$ is the inverse of the atmospheric scale height, $v$ is the velocity of the spacecraft relative to the atmosphere, $\gamma$ the flight-path angle, and $g$ the gravitational acceleration. According to the assumptions made by Vinh, the gravity term should be replaced by an inverse square term, and the atmospheric density varies with an exponential law, at least locally.  




Having defined the adimensional variables in \cref{eq:chapman_var_z,eq:chapman_var_u}, we can follow the procedure outlined by Vinh to compute the overshoot boundary \citep{Hicks2009}:

\bigbreak
\textit{Step 1}:
Select a value of $Z$, namely $Z_{*}$, that is the point of \emph{critical} deceleration. This corresponds to the point where $(a_{decel} / g_0)_{*} = f$. This is the minimum deceleration required by the spacecraft to re-enter in a single passage.

\bigbreak
\textit{Step 2}:
Simultaneously solve the following equations

\begin{align}
	2 \sqrt{\overline{\beta r}} Z_{*} + (\overline{\beta r} - 1) \sin{\gamma_{*}} + \frac{ 2 \sin{\gamma_{*}} \cos{\gamma_{*}}^2 }{ u_{*} } &= 0   \label{eq:overshoot_critical_1} \\
	\frac{ Z_{*} u_{*} \sqrt{\overline{\beta r}} }{ \cos{\gamma_{*}}^2 } \sqrt{1 + \bigg( \frac{C_L}{C_D} \bigg)^2} &= f  \label{eq:overshoot_critical_2}  
\end{align}

to find $u_{*}$ and $\gamma_{*}$. The point defined by $Z_{*}$, $u_{*}$ and $\gamma_{*}$ is a point of minimum deceleration along the trajectory. In \cref{eq:overshoot_critical_1,eq:overshoot_critical_2}, $C_L$ is the lift coefficient and $\overline{\beta r}$ is an average value characteristic of the considered atmosphere.

\bigbreak
\textit{Step 3}:
Solve the two-point boundary value problem defined by the following differential equations

\begin{align}
	\frac{dZ}{ds} &= - \overline{\beta r} Z \tan \gamma  \label{eq:eom_1}  \\
	\frac{du}{ds} &= - \frac{ 2 Z u \sqrt{\overline{\beta r}} }{ \cos \gamma } \Bigg( 1 + \frac{C_L}{C_D} \cos{\sigma} \tan{\gamma} + \frac{\sin \gamma}{2 Z \sqrt{\overline{\beta r}}} \Bigg)  \label{eq:eom_2}  \\
	\frac{d\gamma}{ds} &=  \frac{ Z \sqrt{\overline{\beta r}} }{ \cos \gamma } \Bigg[ \frac{C_L}{C_D} \cos{\sigma} + \frac{\cos \gamma}{Z \sqrt{\overline{\beta r}}} \bigg( 1 + \frac{\cos^2 \gamma}{u} \bigg) \Bigg]  \label{eq:eom_3}
\end{align}

subject to the following boundary conditions

\begin{align}
	Z = Z_{*}, \: u = u_{*}, \: \gamma = \gamma_{*} \quad & \text{at} \quad s = s_{*}  \label{eq:tpbvp_1}  \\
	\frac{ Z_{e} u_{e} \sqrt{\overline{\beta r}} }{ \cos{\gamma_{e}}^2 } \sqrt{1 + \bigg( \frac{C_L}{C_D} \bigg)^2} = f \quad & \text{at} \quad s = 0  \label{eq:tpbvp_2},
\end{align}

until $(a_{decel} / g_0) = f$ again. With this procedure, the entry conditions $Z_{e}$, $u_{e}$ and $\gamma_{e}$ can be obtained. In \cref{eq:eom_1,eq:eom_2,eq:eom_3}, the variable $\sigma$ represents the bank angle, which is assumed to be constant during the motion.

\bigbreak
\textit{Step 4}:
Simultaneously solve the following equations

\begin{align}
	& \frac{u_e^2}{\cos{\gamma_e}^2} - 2 u_e = u_p^2 - 2 u_p \label{eq:overshoot_point_1} \\
	& Z_p = Z_e \sqrt{\frac{u_e}{u_p}} \exp{ \bigg[ \overline{\beta r} \bigg( 1 - \frac{u_e}{u_p} \bigg) \bigg] } \label{eq:overshoot_point_2}
\end{align}

to find $Z_p$, which is the value of $Z$ corresponding to the overshoot boundary. In \cref{eq:overshoot_point_1,eq:overshoot_point_2}, the subscript "$e$" corresponds to the \emph{entry} conditions, while the subscript "$p$" corresponds to the \emph{pericentre} conditions.

\bigbreak
\textit{Step 5}:
Repeat points 1 to 4 with a different value of $Z_{*}$.

\bigbreak
Following steps 1 through 5, the overshoot boundary can be obtained by varying the value of $Z_{*}$. Vinh suggests a range of values for $Z_{*}$ between $1.68 \times 10^{-3}$ and $1.91 \times 10^{-3}$ \citep{Vinh1980,Hicks2009}. \cref{fig:overshootboundary} shows an example of overshoot boundary, considering a ballistic entry ($C_L / C_D = 0$) into Earth atmosphere ($\overline{\beta r} = 900$). To predict if a spacecraft re-enters using the overshoot boundary, we compute the values of $Z_p$ and $u_e$ for the considered trajectory; if this point lies below the overshoot boundary, the spacecraft is then predicted to re-enter.

\begin{figure}[!tbh]
	\centering
	\includegraphics[width=3.25in]{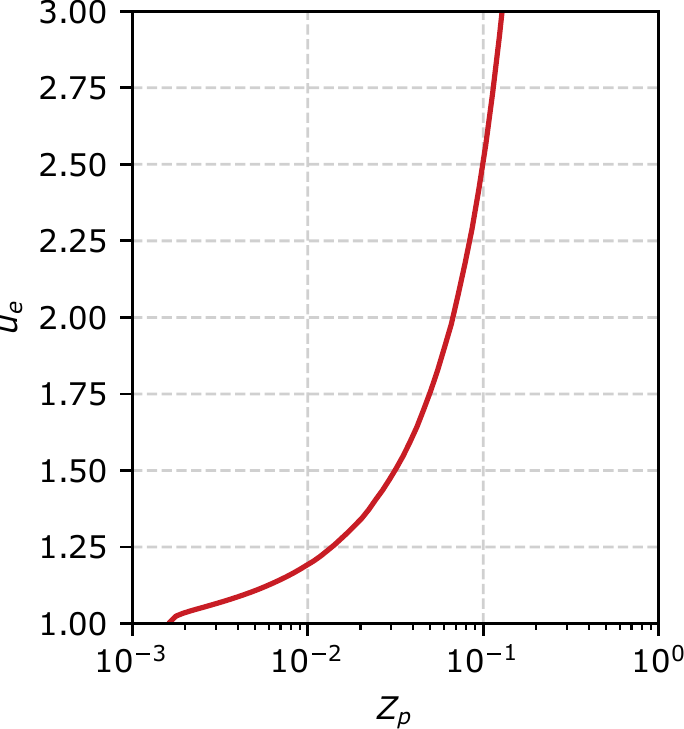}
	\caption{Example of overshoot boundary.}
	\label{fig:overshootboundary}
\end{figure}

It is interesting to observe how the computation of the overshoot boundary using the modified Chapman's variables is independent of the properties of the spacecraft, such as the mass and the cross-section. These characteristics become then important when the computation of the overshoot pericentre radius and the entry velocity need to be computed by inverting \cref{eq:chapman_var_u,eq:chapman_var_z}. Analogous curves to \cref{fig:overshootboundary} can be obtained for other variables of interest, such as $Z_e$ and $\gamma_e$. These expressions can be used in practical applications of the overshoot boundary (see \cref{sec:overshoot_application}).


\section{Re-entry analysis of GSO resonant trajectories}
\label{sec:reentry_analysis}

In this section, the methodologies outlined in \cref{subsec:long-term,subsec:reentry} are applied to the re-entry prediction from GSO resonant trajectories. For such an analysis, a subset of 130 initial conditions has been selected from the disposal maps in \cite{Gkolias2019GEO}. The selection focuses on inclinations greater than 65$^\circ$ and on selected values of the initial eccentricity. \cref{fig:geo_initstates} shows the eccentricity and inclination of the selected initial conditions. The semimajor axis always satisfies the geosynchronous condition, i.e. $a = 42164 \; \si{\kilo\meter}$.

\begin{figure}[!htb]
	\centering
	\includegraphics[width=3.25in]{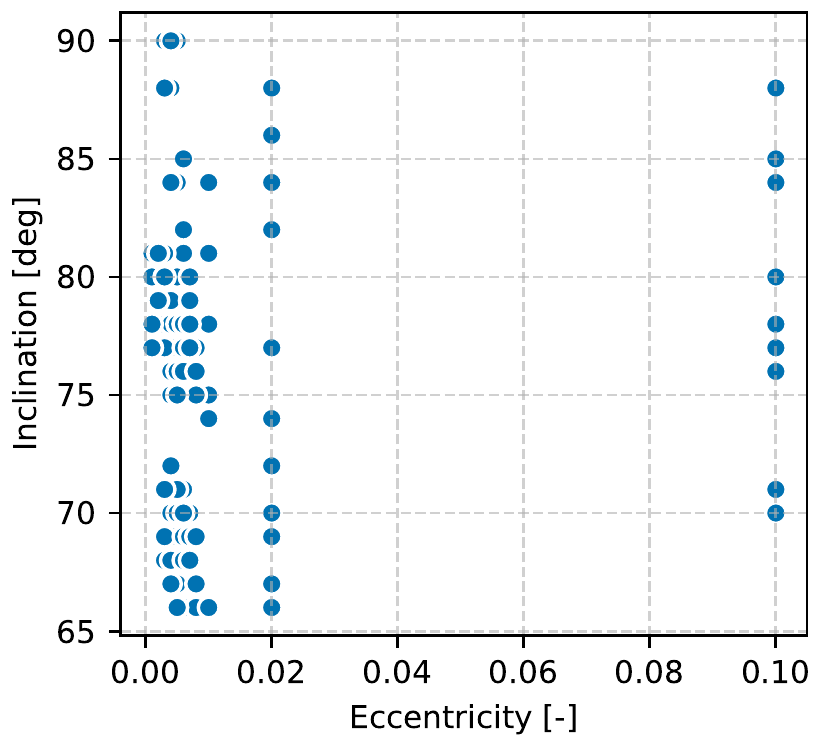}
	\caption{Initial conditions of the selected GSO resonant trajectories.}
	\label{fig:geo_initstates}
\end{figure}

The selected initial conditions are propagated using PlanODyn, including drag, for 150 years or until a re-entry condition is met. In this phase, a \emph{re-entry condition} based on the minimum pericentre altitude has been maintained, setting the threshold to 60 \si{\kilo\meter}. As discussed in \cref{subsec:disposal-maps}, this threshold is somewhat arbitrary and might not predict the re-entry of the satellite at the right conditions. To avoid this ambiguity, for each revolution on the disposal trajectory with a mean pericentre altitude below 120 \si{\kilo\meter}, the \phoenix tool is used to check the re-entry of the spacecraft. To perform the analysis with the \phoenix tool, we need the initial conditions at a specified entry interface. An altitude of 120 \si{\kilo\meter} has been selected for this purpose. To obtain the re-entry conditions at the interface from the long term propagation, a simplified procedure is used. First, we maintain only the long-term disposal trajectories with a mean pericentre altitude below 120 \si{\kilo\meter}. Then, for each orbit revolution, we select only the last orbital parameters before the pericentre passage so that they are the closest to the 120 \si{\kilo\meter} entry interface. Once the orbital parameters are selected, we can find the true anomaly at the entry interface as follows:

\begin{equation}  \label{eq:get_theta}
	\theta_{e, \rm 120} = \arccos{ \bigg[ \frac{1}{e} \bigg( \frac{a}{r_{e_{\rm 120}}} (1 - e^2)  - 1 \bigg) \bigg] },
\end{equation}

where $r_{e_{\rm 120}} = R_{\earth} + 120 \; \si{\kilo\meter}$ is the radius of the entry interface, and $a$ and $e$ are the semi-major axis and eccentricity of the selected orbit, respectively. Once the orbital parameters at the entry interface ($a_{e}$, $e_{e}$, $i_{e}$, $\Omega_{e}$, $\omega_{e}$, $\theta_{e_{\rm 120}}$) are obtained, they can be passed to the \phoenix tool to perform the re-entry simulation and check if the spacecraft is actually predicted to re-enter. This procedure is repeated for each orbit revolution. To perform the simulations with the \phoenix tool, it is also necessary to specify the characteristics of the satellite. For this work, a Beidou-like satellite is selected; its main characteristics are summarised in \cref{tab:spacecraft} and have been obtained as average values from the ESA DISCOS database \citep{DISCOSweb}.

\begin{table}[!htb]
	\centering
	\caption{Main characteristics of the spacecraft configuration used in the analyses.}
	\begin{tabular}{lc}
		\hline
		\textbf{Property} & \textbf{Value} \\
		\hline
		Dry mass (\si{\kilo\gram}) & 1200 \\
		Size (\si{\meter}) & 3 $\times$ 1.7 $\times$ 1.7 \\
		Array area (\si{\meter\square}) & 30 \\
		Average area-to-mass ratio & 0.012 \\
		\hline
	\end{tabular}
	\label{tab:spacecraft}
\end{table}

\cref{fig:phoenix_reentry_prediction} show an example of the results obtained with this procedure. The black line represents the evolution of the entry velocity, $v_e$, (top graph) and entry flight-path angle, $\gamma_e$, (bottom graph) at 120 \si{\kilo\meter} as a function of the pericentre altitude. On top of the black line, the crosses identify skip entries, while the circle represents the first condition that leads to a re-entry. Each marker identifies the re-entry condition for a revolution on the disposal trajectory. In \emph{Phoenix}, a 400 km upper altitude limit is present. Therefore, a skip entry is recorded when a component, after break-up, increases its altitude up to this threshold.

\begin{figure}[!htb]
	\centering
	\includegraphics[width=3.25in]{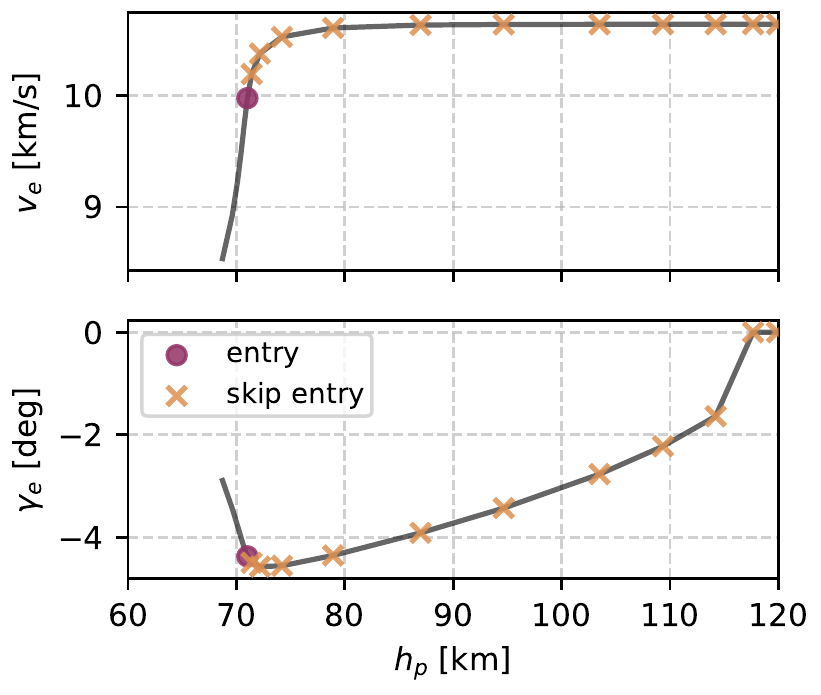}
	\caption{Evolution of the entry velocity (top) and flight-path angle (bottom) as a function of the pericentre altitude. Crosses indicate revolutions for which the \phoenix tool predicted a skip entry, while the circle is the first condition leading to re-entry.}
	\label{fig:phoenix_reentry_prediction}
\end{figure}

If we repeat this analysis on each one of the 130 selected disposal trajectories, we can have a better understanding of the re-entry conditions associated to disposals from GSO resonant trajectories. \cref{fig:hist_reentry} shows a summary of relevant statistics for the entry conditions (corresponding to the circle in \cref{fig:phoenix_reentry_prediction}) resulting from the 130 selected trajectories.

\begin{figure}[htb!]
	\centering
	\begin{subfigure}[b]{0.45\textwidth}
		\centering
		\includegraphics[width=3.25in]{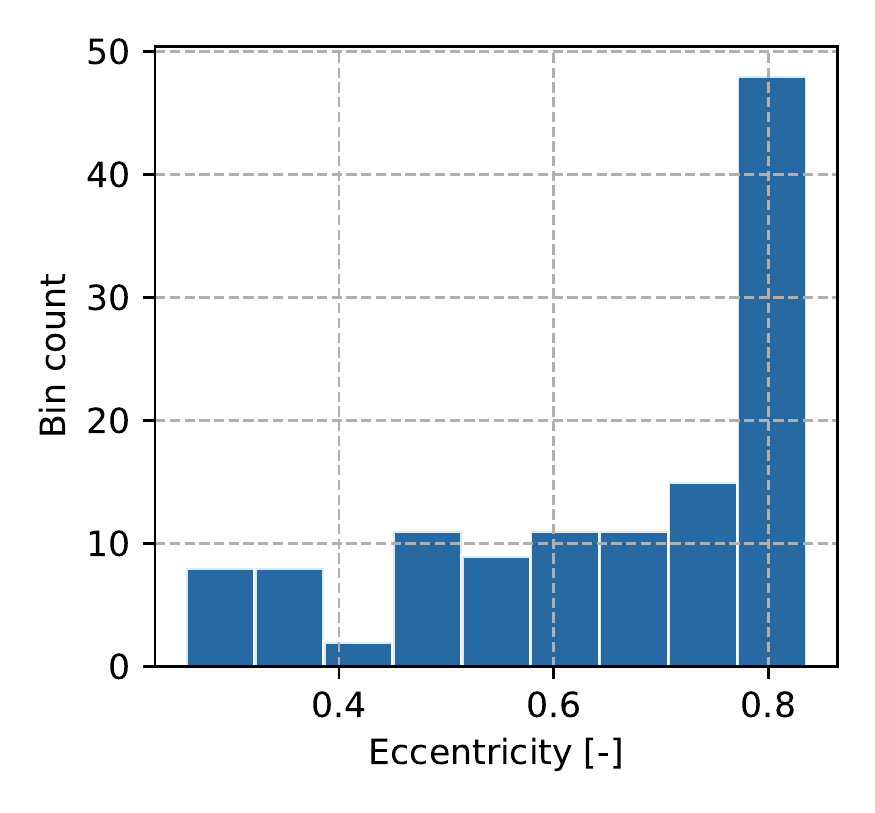}
		\caption{Eccentricity of the orbit at the entry condition.}
		\label{fig:hist_ecc}
	\end{subfigure}
	\hfill
	\begin{subfigure}[b]{0.45\textwidth}
		\centering
		\includegraphics[width=3.25in]{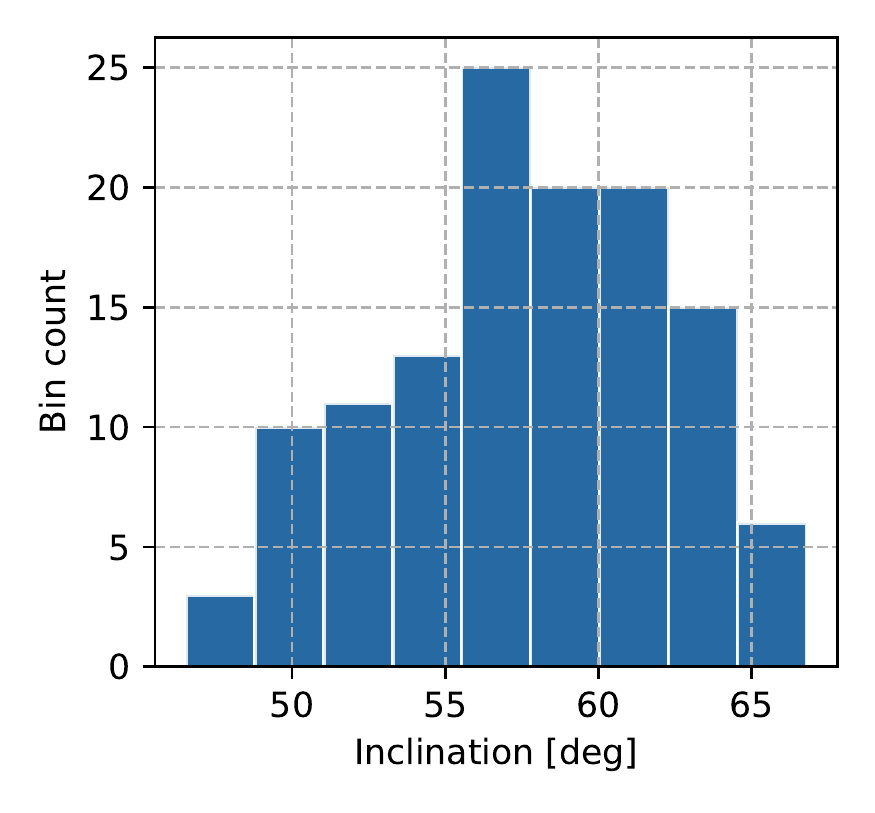}
		\caption{Inclination of the orbit at the entry condition.}
		\label{fig:hist_inc}
	\end{subfigure}
	\hfill
	\begin{subfigure}[b]{0.45\textwidth}
		\centering
		\includegraphics[width=3.25in]{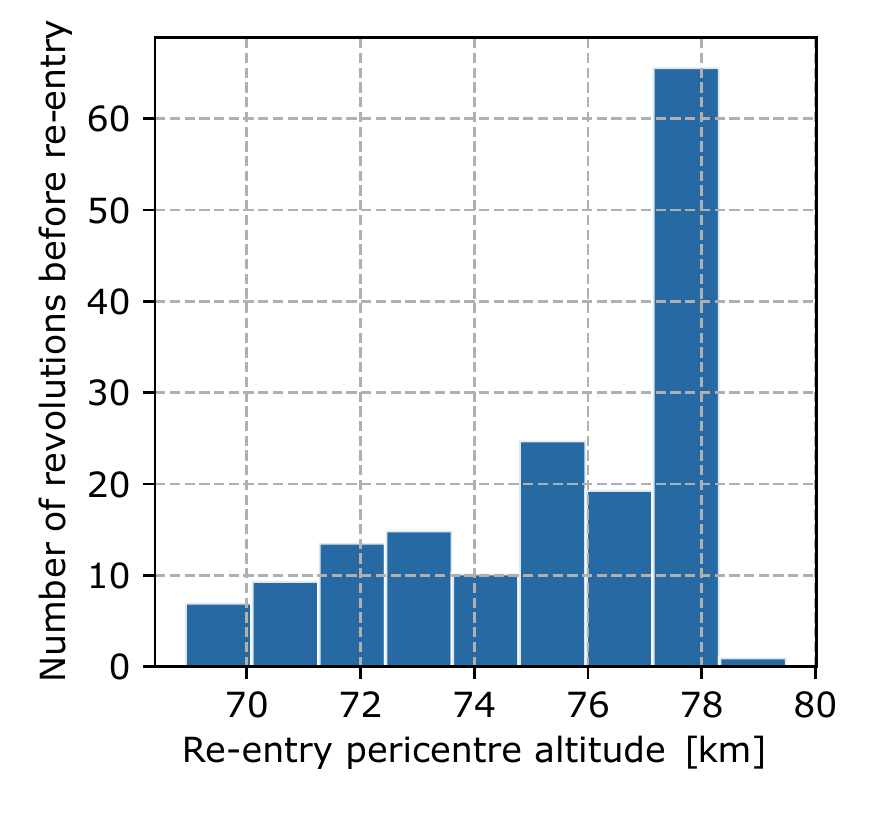}
		\caption{Number of orbit revolutions with pericentre altitude below 120 \si{\kilo\meter} before reaching the entry conditions.}
		\label{fig:hist_norbits}
	\end{subfigure}
	\caption{Orbits characteristics at entry interface.}
	\label{fig:hist_reentry}
\end{figure}

\cref{fig:hist_ecc} shows the distribution of the eccentricity of the orbit at re-entry. For most disposal trajectories (more than one-third) the eccentricity remains high (around 0.8). In 85\% of the cases, the eccentricity is higher than 0.4 and never drops below 0.258. Therefore, we observe how the circularisation effect of \cref{fig:reentry_drag} predicted by PlanODyn does not occur, and the propagation should switch from long-term based to re-entry based at the appropriate moment. \cref{fig:hist_inc} instead shows the distribution of the orbit inclination at re-entry. The initial inclination ranged from 65$^\circ$ to 90$^\circ$ and evolved to a narrower range, between 46$^\circ$ and 67$^\circ$. This evolution of the orbit inclination is interesting, and is also relevant when the casualty risk is considered. In fact, the casualty risk of uncontrolled re-entries is influenced by the latitude band affected by the event, which in turn is connected to the inclination of the orbit. Finally, \cref{fig:hist_norbits} shows the average number of revolutions performed by the spacecraft on the disposal orbit before meeting the re-entry conditions, for orbits with a pericentre altitude below 120 \si{\kilo\meter}. The number of revolutions is presented as a function of the pericentre altitude of the re-entry orbit. The analysis shows that several passages through the atmosphere may be needed to reach a re-entry condition, which may generate high enough aerothermodynamics loads to result in fragmentations or detachment of components from the spacecraft before re-entry. On one hand, this occurrence may release unwanted and dangerous debris into orbit, resulting in potential threats to other satellites, if the detached components skip the Earth's atmosphere. On the other hand, this phase of catastrophic decay may have potential benefits in reducing the casualty risk. In fact, a partial disassembly of the spacecraft can expose the internal components to early re-entry heating, and the possible decrease in entry speed and flight-path angle could generate higher heat loads, thus improving the demise. \cref{fig:hist_norbits} also shows that a higher number of revolutions is associated with orbits with higher pericentre altitudes and lower eccentricities. As increasing the number of passages through the atmosphere can increase the chance of early breakups, this circularisation phenomenon should be avoided, if the risk of releasing unwanted debris into orbit must be minimised.

\section{Re-entry predictions with the overshoot boundary}
\label{sec:overshoot_application}

As outlined in \cref{sec:reentry_analysis}, understanding what are the conditions leading to the final re-entry of the spacecraft required switching from the long-term propagation to the destructive re-entry analysis. This switching is necessary when performing a full demisability and casualty risk analysis. However, it might be a cumbersome procedure when the intention is limited to discerning between the entry/skip behaviour, such as in the design and optimisation of the disposal trajectory and limit the number of low-altitude passages. To tackle this problem, we introduced the concept of \emph{overshoot boundary} in \cref{subsec:overshoot}. Since the overshoot boundary can be pre-computed as in \cref{fig:overshootboundary}, it is a convenient tool for a quick assessment of re-entry predictions. In this section, the application of the overshoot boundary is specialised to the case in exam of GSO resonant re-entries, and the predictions obtained with it are compared with the ones of the \phoenix tool.

To apply the overshoot boundary to the case in exam, it is first necessary to specify the characteristics of the spacecraft and the assumption made during the analysis. For the pre-computation of the adimensional overshoot boundary (\cref{subsec:overshoot}), it is necessary to specify an average property of the atmosphere via the $\overline{\beta r}$ parameter, and the lift-to-drag ratio. For the case in exam, a value of $\overline{\beta r} = 900$ \citep{Hicks2009} and a $C_L / C_D = 0$ have been used. The latter assumption supposes a random tumbling re-entry, which produces a negligible lift force. As we are analysing uncontrolled re-entries this is considered a suitable assumption. In addition, this choice matches the dynamics used in the \phoenix tool, which assumes a random tumbling motion of the spacecraft.

At this point, we need to specify the cross-section, $S$, and drag coefficient, $C_D$, of the spacecraft, as well as the characteristics of the atmosphere. Since the overshoot boundary is based on a strictly exponential atmosphere of the form $\rho (h) = \rho_0 e^{-\beta h}$, it is necessary to specify the value of $\rho_0$ and $\beta$ . For the case in exam, $\rho_0 = 1.225 \; \si{\kilo\gram \per \meter\cubed}$, which is the standard sea-level value, and $\beta = 1.4544 \times 10^{-4} \; \si{\per \meter}$. The value of $\beta$ has been obtained averaging the piecewise exponential atmosphere described in \cite{Vallado2013}.
As previously mentioned, we assume a random tumbling motion; therefore, for the cross-section $S$, we use an average value including the solar panels area as follows:

\begin{equation}  \label{eq:avg_cross_section}
	S = \overline{S} = \frac{1}{3} \left(A_{sp} + W H + L H + L W\right).
\end{equation}

where $L$, $W$ and $H$ are the length, width and height of the main body of the spacecraft, and $A_{sp}$ is the solar panels area. The characteristics of the spacecraft are the same as the ones presented in \cref{tab:spacecraft}. 

For the drag coefficient, we are focusing on the parts of the disposal trajectory with a pericentre altitude between 60 and 120 \si{\kilo\meter}. In this altitude region, the spacecraft can encounter all the three different flow regimes: free-molecular, transition, and continuum. Therefore, to better capture the dynamical behaviour, a variable drag coefficient has been used. We use the Knudsen number, $\textit{Kn} = l_c / \lambda$, to discriminate between these three flow regimes, where $l_c$ and $\lambda$ are the characteristic length of the spacecraft and the atmospheric mean-free path, respectively. Specifically, for $\mathit{Kn} < 0.01$ the flow is classified as continuum, while for $\mathit{Kn} > 1$ the flow is classified as free-molecular. Values in between ($0.01 \leq \mathit{Kn} \leq 1$) are associated to transition flows. The drag coefficient depends on the type of flow encountered; therefore, different expressions must be used for the three flow regimes. Specifically, \cref{eq:drag_fm,eq:drag_c} represent the expressions for the free-molecular and continuum drag coefficients, respectively \citep{Klett1964,TLC18Sensitivity,Trisolini2018_AESCTE}.

\begin{align}
	\mathit{CD}_{\mathrm{fm}} &= 1.57 + 0.785 \cdot \frac{W}{L}  \label{eq:drag_c} \\
	\mathit{CD}_{\rm c} &= 1.83 \cdot \left(0.393 + 0.178 \cdot \frac{W}{L}\right) \label{eq:drag_fm}
\end{align}

For the transition regime, a bridging function is used that connects the values of the continuum and free-molecular drag coefficients as a function of the Knudsen number, as follows:

\begin{equation}  \label{eq:bridging}
	\mathit{CD}_{\rm t} (\mathit{Kn}) = \mathit{CD}_c + (\mathit{CD}_{\mathrm{fm}} - \mathit{CD}_{\rm c}) \cdot \Big( \sin \left( \pi \left(0.5 + 0.25 \log_{10} \textit{Kn}\right) \right) \Big)^3
\end{equation}

\cref{eq:bridging} is the same bridging function used in the \phoenix tool for the best compatibility between the analyses. The drag coefficient used in the computation of the overshoot boundary is then obtained from \cref{eq:drag_c,eq:drag_fm,eq:bridging} using the value of the Knudsen number corresponding to the pericentre altitude. The relation between the Knudsen number and the altitude is derived from the 1976 U.S. Standard Atmosphere \citep{1976_std_atm} by extracting the relation between the mean-free path and the altitude, and choosing the value of the characteristic length ($l_c = L$ for the case in exam). The choice of the 1976 U.S. Standard Atmosphere maintains the compatibility with the \phoenix tool as it is based on the same atmosphere model.


%

Once selected the parameters describing the disposal, we can check the re-entry of the spacecraft using the overshoot boundary with the following procedure:

\begin{enumerate}
	\item For each revolution on the disposal orbit, compute the pericentre radius, $r_p$, (in the same fashion as described in \cref{sec:reentry_analysis}).
	\item Compute the value of $Z_p$.
	\item Use $Z_p$ to compute the adimensional velocity belonging to the overshoot boundary ($u_{e, b}$) and the entry interface adimensional radius, $Z_e$.
	\item Convert $Z_e$ to $r_e$ inverting \cref{eq:chapman_var_z}.
	\item Use $r_e$ to compute the velocity ($v_{e,d}$) and flight-path angle ($\gamma_{e,d}$) along the disposal trajectory at the entry interface as follows:
	
	\begin{align}
		v_{e,d} &= \sqrt{ \mu_{\earth} \bigg( \frac{2}{r_e} - \frac{1}{a_d} \bigg) } \\
		\gamma_{e,d} &= \arctan{ \bigg( \frac{e_d \sin \theta_{e, d}}{1 + e_d \cos \theta_{e, d}} \bigg) },
	\end{align}
	
	where $a_d$ is the semi-major axis of the disposal orbit, $e_d$ its eccentricity, and $\theta_{e,d}$ is the true anomaly at the entry interface along the disposal orbit, and can be computed with \cref{eq:get_theta}.
	\item Compute the adimensional velocity at entry along the disposal orbit, $u_{e,d}$, with \cref{eq:chapman_var_z,eq:chapman_var_u}.
	
\end{enumerate}

At the end of the procedure, if $u_{e,d}$ is smaller $u_{e,b}$ the spacecraft is predicted to re-entry. If we repeat points 1 through 6 for each revolution, we obtain the evolution of the entry conditions compared to the overshoot boundary. \cref{fig:overshoot_comparison} summarises the results of such a procedure for all the 130 selected disposal trajectories. The red line represents the overshoot boundary of the case in exam as a function of the pericentre altitude, $h_p$. The blue shaded lines represent the evolution of the entry conditions for each disposal trajectory. Following a line from right to left, the entry conditions evolve, passing from the region above the boundary to the region below. When this occurs, the spacecraft is predicted to re-enter. The different shades of blue represent the evolution of the orbit eccentricity. As we move from right to left, after each revolution, the eccentricity decreases. For the initial GSO orbits of \cref{fig:geo_initstates}, the overshoot boundary predicts the re-entry for disposal orbits with a pericentre altitude between 70 and 80 \si{\kilo\meter}, which is in agreement with the results of the \phoenix tool (\cref{fig:hist_norbits}). Also in accordance with \cref{fig:hist_ecc}, lower pericentre altitudes are associated with more eccentric re-entries. 

\begin{figure}[htb!]
	\centering
	\includegraphics[width=3.25in]{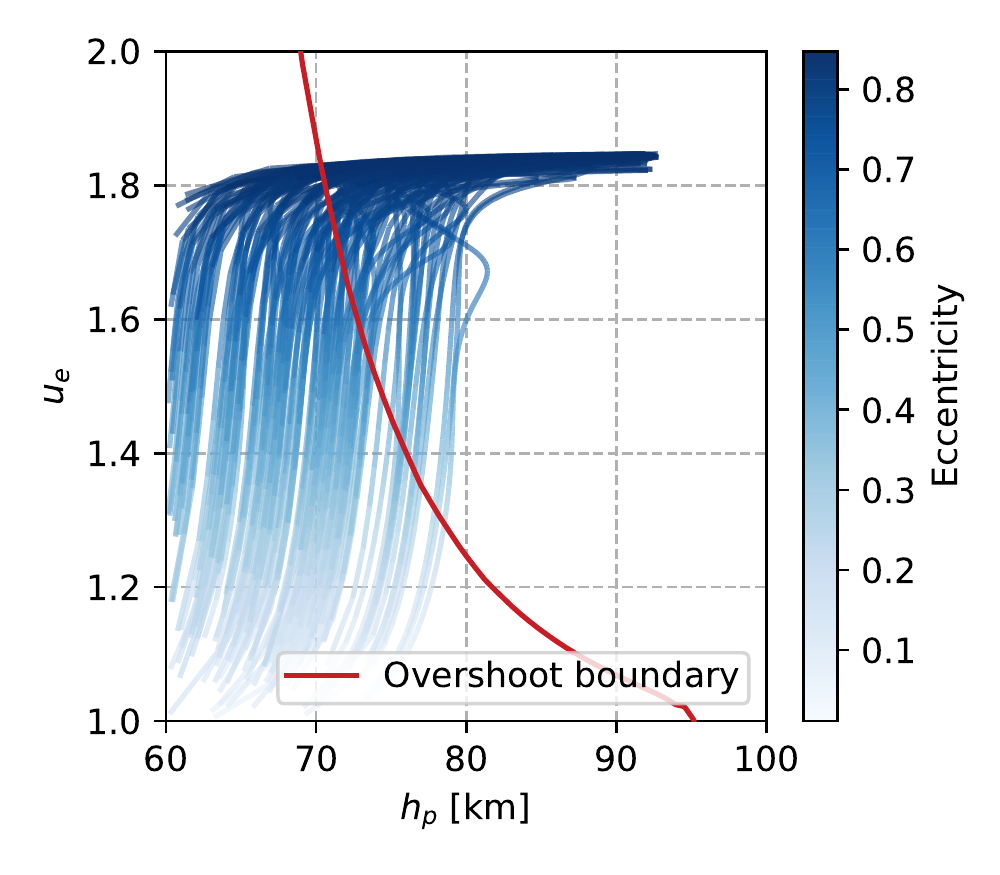}
	\caption{Evolution of the entry conditions as a function of the pericentre altitude of the disposal trajectory. Different shades of blue represent the evolution of the orbit eccentricity. In red, the overshoot boundary.}
	\label{fig:overshoot_comparison}
\end{figure}

At this point, we compare the re-entry predictions of the overshoot boundary to the ones obtained in \cref{sec:reentry_analysis} with the \phoenix tool. \cref{fig:overshoot_comparison_check} shows this comparison. Similarly to \cref{fig:phoenix_reentry_prediction}, the orange crosses identify pericentre passages with skip conditions and the circles the re-entries for a subset of 70 of the 130 disposal trajectories. For better clarity, only the last skip condition and the first entry condition are represented. The conditions predicted by the overshoot boundary are comparable to the results of \emph{Phoenix}. In most cases, the overshoot boundary marginally overpredicts the re-entry conditions of \emph{Phoenix}, with a higher adimensional entry velocity, $u_e$. However, the purpose of the overshoot boundary is not to precisely predict the entry velocity of the satellite but to provide an analytical method to discriminate between the entry / no-entry behaviour. As demonstrated by \cref{fig:overshoot_comparison_check}, the overshoot boundary can be used as an analytical method to predict the re-entry of satellites from GSO, giving comparable results to destructive re-entry codes. It can be used for the preliminary design of disposal trajectories and the optimisation of disposal manoeuvres to improve their reliability.

\begin{figure}[htb!]
	\centering
	\includegraphics[width=3.25in]{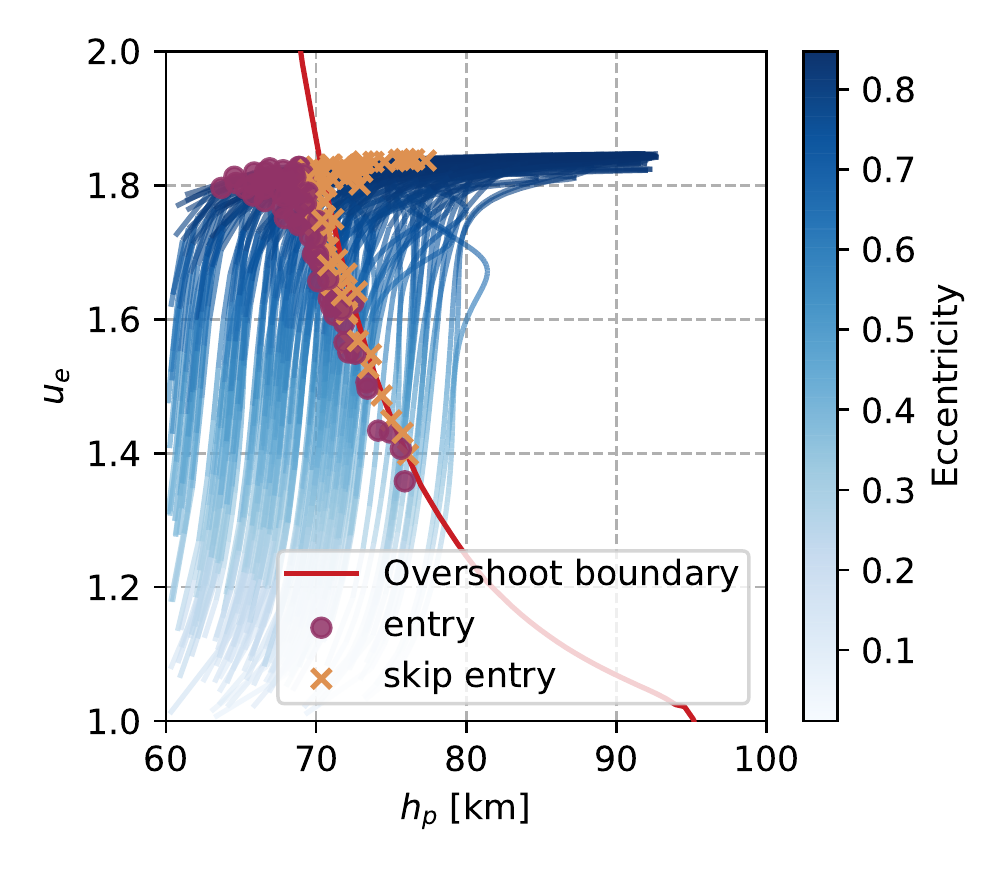}
	\caption{Evolution of the entry conditions as a function of the pericentre altitude of the disposal trajectory. In red the overshoot boundary. Crosses represent the last skip entry, while circles the first entry condition predicted by \emph{Phoenix}.}
	\label{fig:overshoot_comparison_check}
\end{figure}

\section{Demisability analysis}
\label{sec:demisability}

Even though the overshoot boundary gives a simpler way of predicting the re-entry, it is still important to understand if the spacecraft poses a threat to people and properties on the ground. Using the \emph{Phoenix} destructive re-entry tool, we can perform a demisability analysis. The results of \cref{sec:reentry_analysis} show the main characteristics of the entry conditions of GSO disposal orbits. We can convert the orbital parameters at the 120 \si{\kilo\meter} entry interface ($a_{e}$, $e_{e}$, $i_{e}$, $\Omega_{e}$, $\omega_{e}$, $\theta_{e}$) into flight variables (longitude ($\lambda_e$), latitude ($\varphi_e$), radius ($r_e$), velocity ($v_e$), flight-path angle ($\gamma_e$), and heading angle ($\chi_e$)). \cref{fig:pairplot_reentry} shows the pair plot of the re-entry conditions for the latitude, velocity and flight-path angle. The subplots along the main diagonal represent the one-dimensional marginals of the variables identified by the bottom labels. The off-diagonal subplots represent the two-dimensional marginals. Lighter shades of blue correspond to a higher number of occurrences in the corresponding bin.

\begin{figure}[htb!]
	\centering
	\includegraphics[width=3.25in]{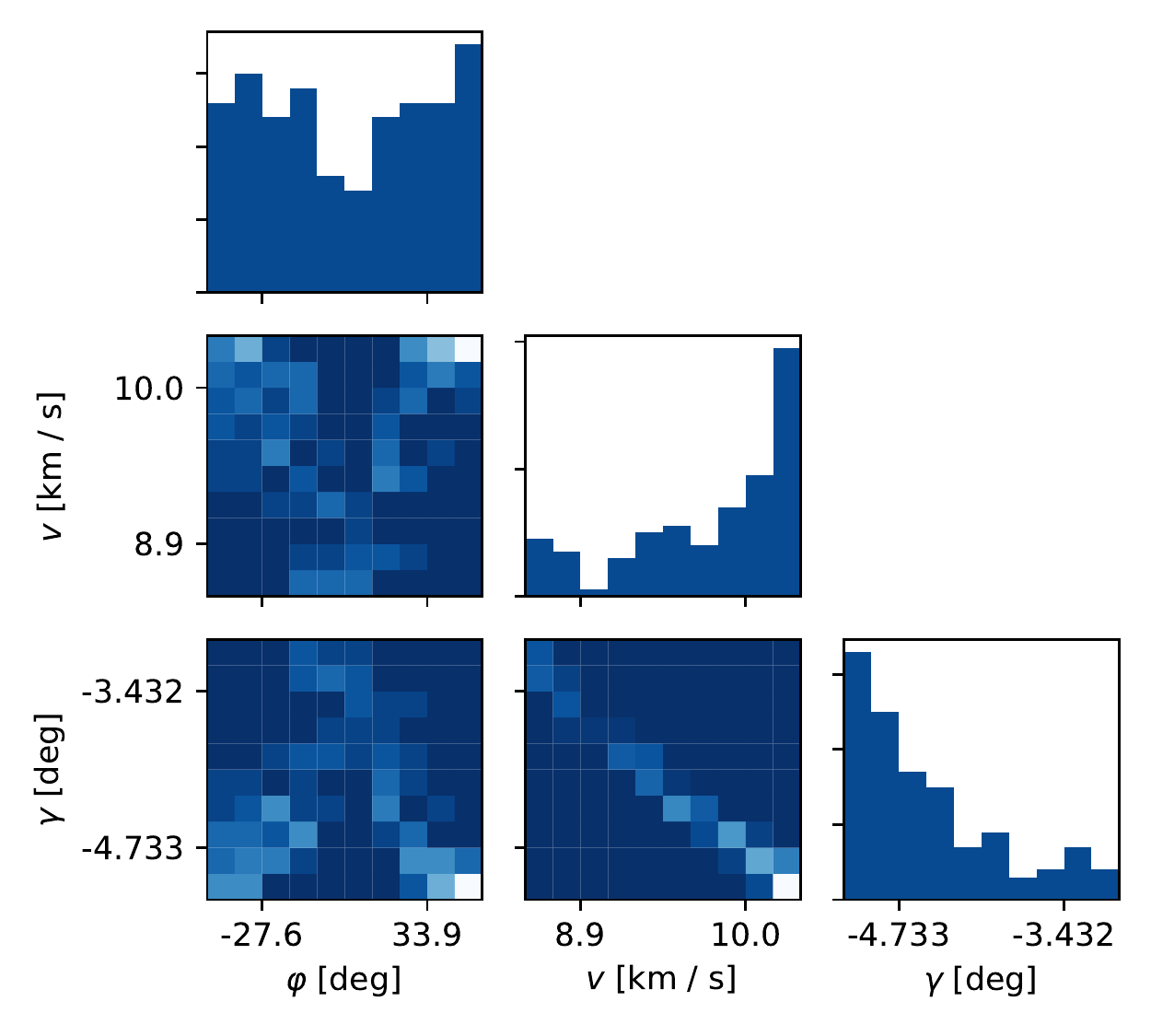}
	\caption{Pairplot of the distribution of the entry conditions predicted by \emph{Phoenix}. Representing the latitude $\varphi$, velocity $v$ and flight-path angle $\gamma$. Lighter shades are associated to a higher number of occurrences in the corresponding bin.}
	\label{fig:pairplot_reentry}
\end{figure}

\cref{fig:pairplot_reentry} shows a prevalence of re-entries with steeper flight-path angles and higher velocities, which is in agreement with \cref{fig:hist_ecc}, which shows a prevalence of re-entry trajectories with high eccentricities. It can also be observed the correlation between the entry velocity and the entry flight-path angle, with higher velocities associated to steeper flight-path angles.

For a better understanding of the risk that GSO satellites can pose, a destructive re-entry analysis has been performed, using a simplified spacecraft configuration (see \cref{sec:appendix_config}). The configuration includes the most critical components in terms of demise, such as reaction wheels assemblies, tanks, batteries and antennas. The configuration of \cref{tab:spacecraft_config} models about 42\% of the 1200 \si{\kilo\gram} of the spacecraft dry mass. This amount is considered sufficient for the demisability analysis as it contains the most critical components. For the demisability analysis, a destructive re-entry simulation using the \phoenix tool has been performed for each one of the entry conditions of \cref{fig:pairplot_reentry}. The breakup altitude has been sampled randomly between 55 and 80 \si{\kilo\meter}. For each initial condition, ten different breakup altitudes have been sampled for a total of 1300 re-entry simulations. \cref{fig:reentry_trajectories} show the trajectory evolution of selected components in terms of altitude vs time after the breakup. The colour scale indicates the flight path angle at the initial instant. Three relevant examples have been selected. \cref{fig:traj_tank} shows the re-entry behaviour of the pressuriser tank. This is a large component made of stainless steel that never demises for all the considered initial conditions. It also has an interesting entry evolution; in fact, about 34.2\% of the entry trajectories skipping behaviour. It is essential to consider this behaviour as massive components may possess enough energy after the catastrophic breakup to skip the atmosphere, given the high entry velocities of GSO re-entry trajectories (\cref{fig:pairplot_reentry}). \cref{fig:traj_battbox} shows the entry trajectories of a battery box. It is a medium-sized component made of aluminium alloy, which, as expected, tends to demise early on most occasions. However, as \cref{fig:traj_battbox} shows, in some cases, even these types of components can survive the re-entry. This is the case for low breakup altitudes and steeper flight-path angles. Finally, \cref{fig:traj_trx} shows the entry trajectories of a transceiver core. For some trajectories, skip behaviour is observed, but the component demises and, in the end, do not exit the atmosphere. Among the remaining trajectories, some reach the ground while others stop early, indicating a quicker demise. Some of them also stop relatively late in the simulation. These last cases indicate simulations for which the energy of the component has dropped below the 15 \si{\joule} threshold.

\begin{figure}[htb!]
	\centering
	\begin{subfigure}[b]{0.48\textwidth}
		\centering
		\includegraphics[width=3.25in]{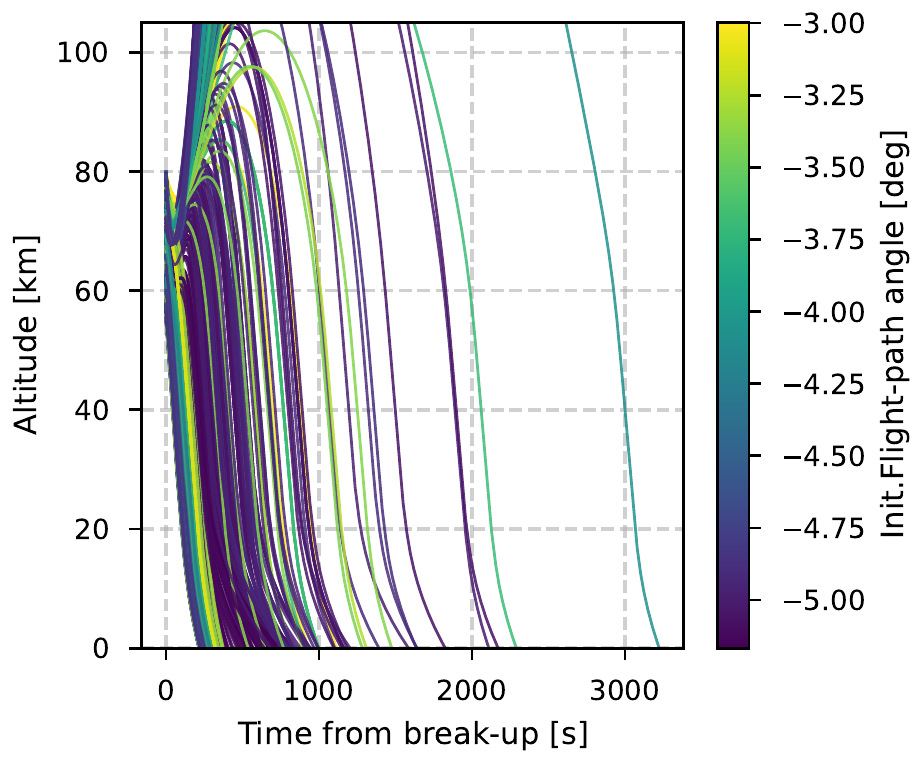}
		\caption{Pressuriser tank.}
		\label{fig:traj_tank}
	\end{subfigure}
	\hfill
	\begin{subfigure}[b]{0.48\textwidth}
		\centering
		\includegraphics[width=3.25in]{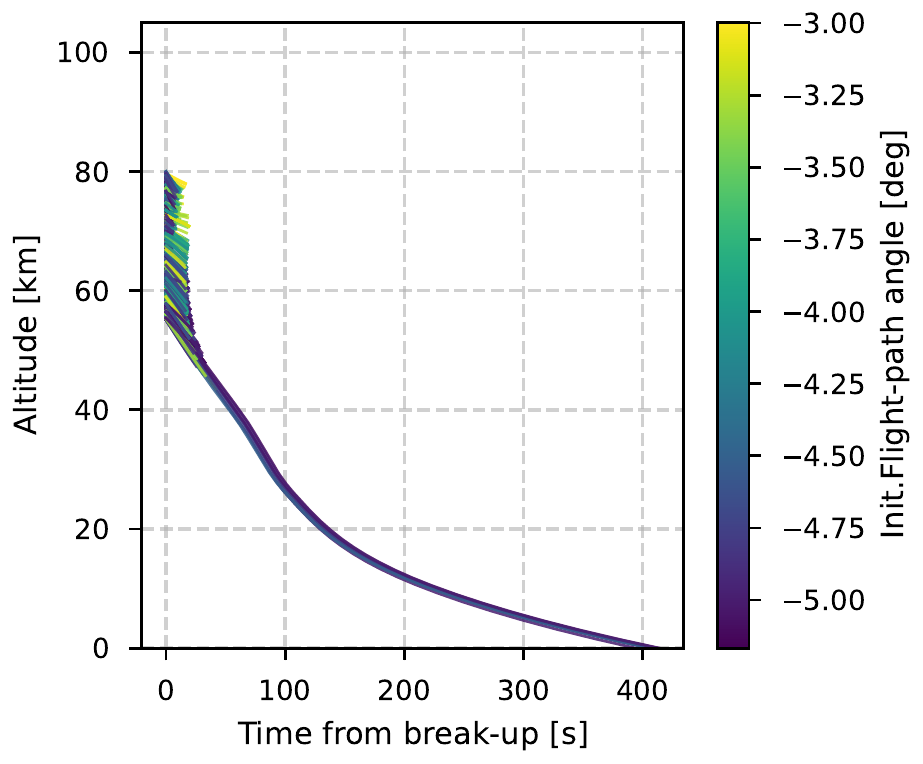}
		\caption{Battery box.}
		\label{fig:traj_battbox}
	\end{subfigure}
	\hfill
	\begin{subfigure}[b]{0.48\textwidth}
		\centering
		\includegraphics[width=3.25in]{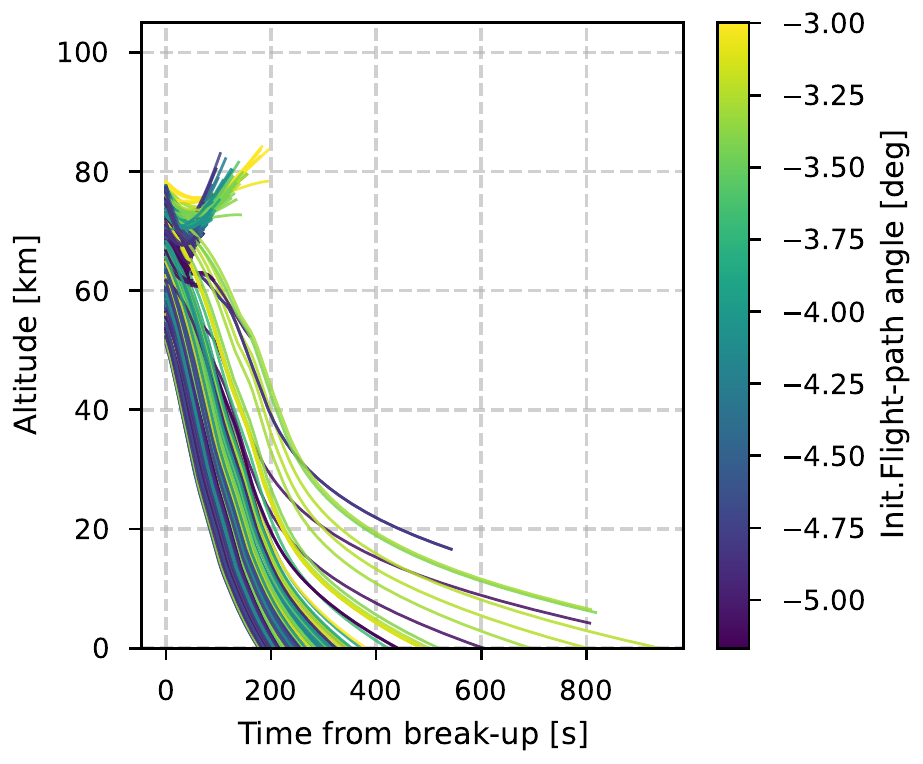}
		\caption{Transceiver core.}
		\label{fig:traj_trx}
	\end{subfigure}
	\caption{Trajectory evolution of the altitude vs time for selected components. a) pressuriser tank, b) battery box, and c) transceiver core. The colormap indicates the initial flight-path angle of the trajectory.}
	\label{fig:reentry_trajectories}
\end{figure}

For each of the propagated trajectories, we also perform a casualty risk analysis. As uncontrolled re-entries have been considered, the casualty expectancy, $E_c$ can be computed with the following expression \citep{ESASpaceDebrisMitigation2015}:

\begin{equation}  \label{eq:casualty_risk}
	E_c = A_c \cdot \rho_p (i, \Delta \varphi, \omega),
\end{equation}

where $A_c$ is the total casualty area due to all the surviving fragments and $\rho_p$ is the latitude-averaged population density, which is a function of the orbit inclination, $i$, the argument of perigee at the epoch of atmospheric capture, $\omega$ (only considered for eccentric orbits), and the latitude resolution, $\Delta \varphi$. The casualty area can be computed as follows:

\begin{equation}  \label{eq:casualty_area}
	A_c = \sum_{j = 1}^{n_c} (\sqrt{A_{j}} + \sqrt{A_h})^2,
\end{equation}

where $n_c$ is the number of survived components, $A_{j}$ is the average projected area of the $j$th survived component, and $A_h = 0.36 \; \si{\meter\squared}$ is the cross-section of a human. The casualty cross-section is computed from the initial cross-section of the components and then removing a uniform thickness proportional to the ablated mass during the re-entry. 
The average population density $\rho_p$ depends on the longitudinal averaged population of the world and on the latitude-dependent impact probability, as follows:

\begin{equation}
	\rho_p (i, \Delta \varphi, \omega) = \sum_{\varphi = -90^\circ}^{\varphi = +90^\circ} \rho (\varphi, \Delta \varphi) \cdot P_i (i, \varphi, \Delta \varphi, \omega),
\end{equation}

where $\rho$ is the longitude-average population density at the latitude $\varphi$, with a $\Delta \varphi$ margin around $\varphi$, and $P_i$ is the impact probability distribution. The expression for the impact probability depends on the type of re-entry, specifically on its eccentricity. As we have seen in \cref{sec:reentry_analysis}, for GSO missions, all re-entries are eccentric. For these types of entries, the debris mitigation guidelines of the European Space Agency (ESA) specify that the entry location can be estimated with sufficient accuracy knowing the argument of pericentre and the revolution number at which the re-entry occurs \citep{ESASpaceDebrisMitigation2015}. However, the current analysis focuses on a general understanding of the demisability of GSO satellites. Therefore, to maintain the generality of the obtained results, we assumed the re-entry is equally probable in the latitude band covered by the disposal orbit at re-entry. The world population used in the analysis is derived from the 2020 GPW database \citep{CIESIN2020} and scaled according to the date, assuming a 1.08\% growth per year \citep{UnitedNations2019population}.

\cref{fig:hist_casualty} shows the distribution of the casualty risk for the 1300 simulated trajectories. It is possible to observe that the casualty risk is always greater than the 10$^{-4}$ threshold, indicating that the representative GSO spacecraft considered is never compliant with the casualty risk regulation, despite we modelled less then 50\% of the spacecraft mass. Therefore, it is probable that most of GSO satellites will not be compliant with the casualty risk regulations and would probably require a controlled or semi-controlled re-entry. An alternative that can be considered in the design stage is to implement design-for-demise techniques on selected components to improve their demisability and the overall casualty footprint of the spacecraft. \cref{fig:hist_casualty} also shows that most occurrences are concentrated in the lower end of the casualty risk range, indicating that there is a core of components usually surviving the re-entry (tanks and reaction wheels). Exploring design-for-demise options for these component could lead to a compliant design in most cases. Additionally, end-of-life strategies for GSO satellites could also contribute to reduce the overpopulation of the graveyard orbit by allowing less satellites to be disposed in these regions.

\begin{figure}[!htb]
	\centering
	\includegraphics[width=3.25in]{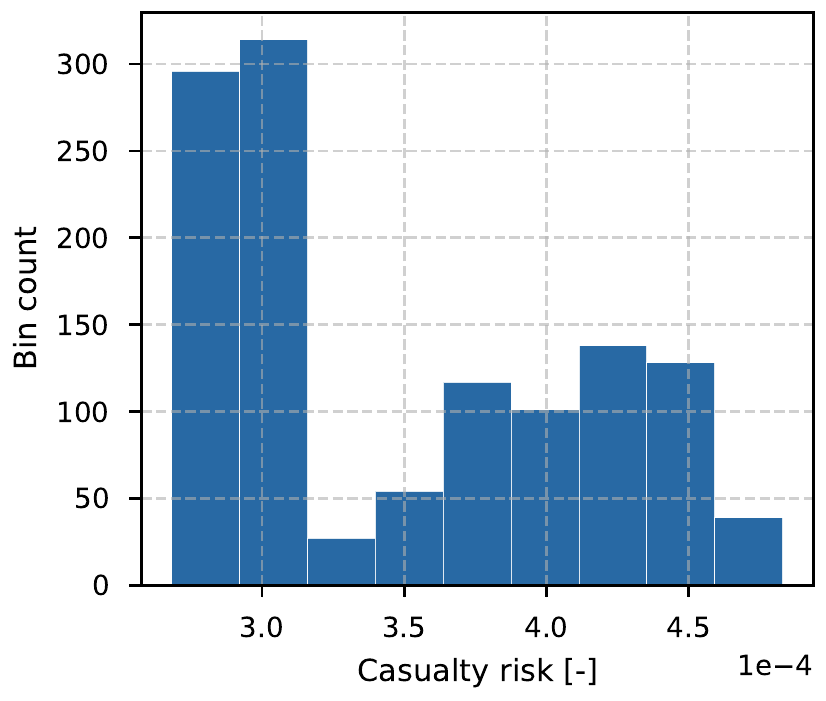}
	\caption{Predicted casualty risk for the selected re-entry trajectories.}
	\label{fig:hist_casualty}
\end{figure}

\subsection{Preliminary analysis of the aerothermodynamic loads before re-entry}
\label{subsec:loads_analysis_pre_reentry}

\cref{sec:reentry_analysis,sec:overshoot_application} showed that the re-entry conditions from GSO resonant trajectories tend to be characterised by low pericentre passages. As shown in \cref{fig:phoenix_reentry_prediction,fig:hist_norbits}, several low-altitude passages occur before the spacecraft is able to re-enter. It is thus interesting to check what are the aerothermal loads encountered by the spacecraft during these passages. Assuming an initial temperature of 300 \si{\kelvin} and a random tumbling motion, using the \phoenix tool, a re-entry simulation has been performed for each atmospheric passage. The initial conditions for the re-entry simulation have been obtained in the same fashion as in \cref{sec:reentry_analysis} and, for each passage, the initial temperature is reset to 300 \si{\kelvin}, thus assuming the heat accumulated during one passage is dispersed before the next one. The parent spacecraft structure has been modelled as an aluminium 6061-T6 box with a 5 \si{\milli\meter} thickness. \cref{fig:temp_evolution,fig:pdyn_evolution} show the evolution of the temperature and dynamic pressure experienced by the parent spacecraft for a selected disposal trajectory, respectively. In both plots, each line represents a pericentre passage as a function of the flight-path angle ($x$-axis) and altitude ($y$-axis). The colours represent the temperature of the spacecraft and the dynamic pressure acting on it. \cref{fig:temp_evolution} shows that the parent spacecraft reaches the maximum temperature (the melting temperature of the considered aluminium alloy) after just a couple of passages in the atmosphere. This is an interesting and important aspect, as the structural integrity of the spacecraft might be compromised before it reaches a re-entry orbit. Of course, this type of fragmentation must be avoided as it could release components and fragments into highly eccentrical orbits around the Earth.

\begin{figure}[!htb]
	\centering
	\includegraphics[width=3.25in]{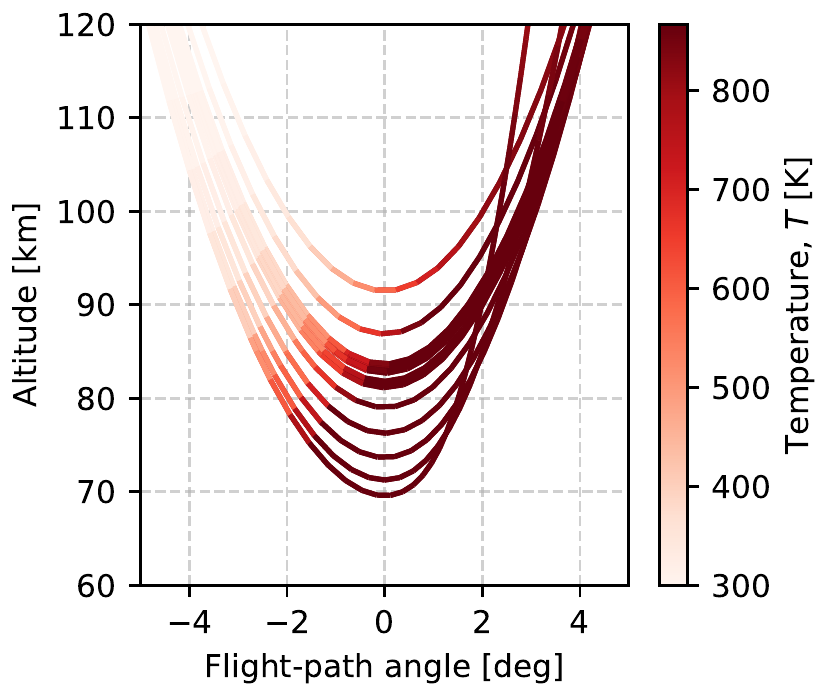}
	\caption{Evolution of the parent spacecraft temperature at each atmospheric passage before re-entry for a selected trajectory.}
	\label{fig:temp_evolution}
\end{figure}

\cref{fig:pdyn_evolution} shows instead that the evolution of the dynamic pressure for the considered trajectory has a peak at about 3500 \si{\pascal}. This value is considerably lower than the yield strength of the considered aluminium alloy, which is 276 \si{\mega\pascal}. Even considering that the yield strength at high temperatures drops below 10 \si{\mega\pascal} \citep{MatWebLLC2015}, the value is still one order of magnitude greater than the maximum dynamical pressure experienced by the spacecraft.

\begin{figure}[!htb]
	\centering
	\includegraphics[width=3.25in]{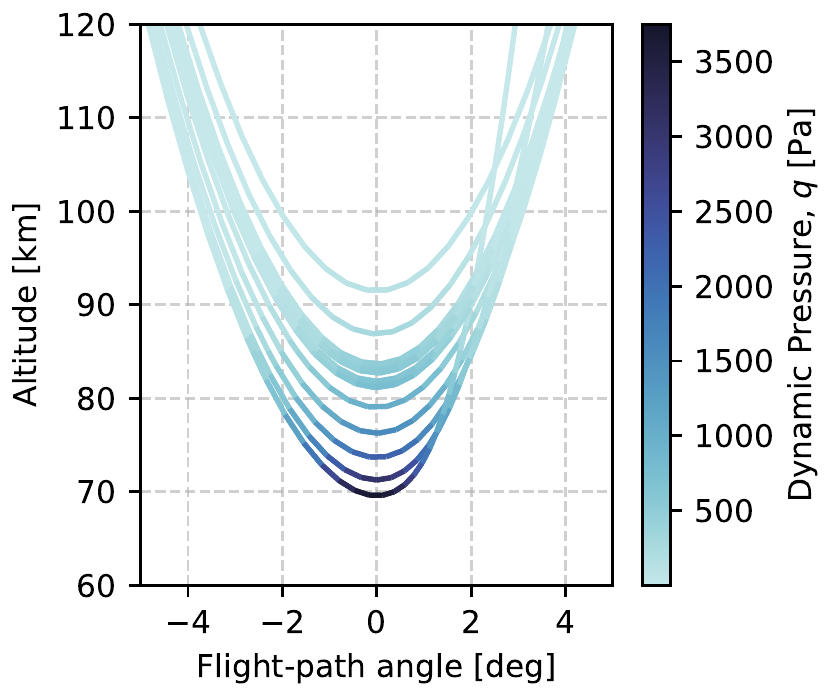}
	\caption{Evolution of the dynamic pressure experienced by the parent spacecraft at each atmospheric passage before re-entry for a selected trajectory.}
	\label{fig:pdyn_evolution}
\end{figure}

Repeating the analysis for all the selected disposal trajectories, we can have a better understanding of the aerothermal loads experienced by the spacecraft. Specifically, \cref{fig:hist_tmelt} shows the distribution of the pericentre altitudes at which the melting temperature ($T_{\rm melt}$) was first reached during the disposal. The figure indicates that for these types of trajectories the melting temperature has been reached for orbits with pericentre altitudes ranging between about 80 to 90 \si{\kilo\meter}, with a prevalence of orbits with higher pericentre altitudes. Given that in several low-altitude passages the parent structure reaches its melting temperature, it is possible that in one of these passages, the additional heating converts the structure from solid to liquid, leading to the break-up of the spacecraft.

\cref{fig:hist_pdyn} shows instead a statistic on the maximum dynamic pressure ($q_{\rm max}$) experienced by the parent spacecraft, considering all the selected disposal trajectories. It can be observed how the dynamic pressure ranges between about 1 and 4.5 \si{\kilo\pascal}, thus maintaining a value below the yield strength of the material in all the considered cases.

\begin{figure}[htb!]
	\centering
	\begin{subfigure}[t]{0.45\textwidth}
		\centering
		\includegraphics[width=3.25in]{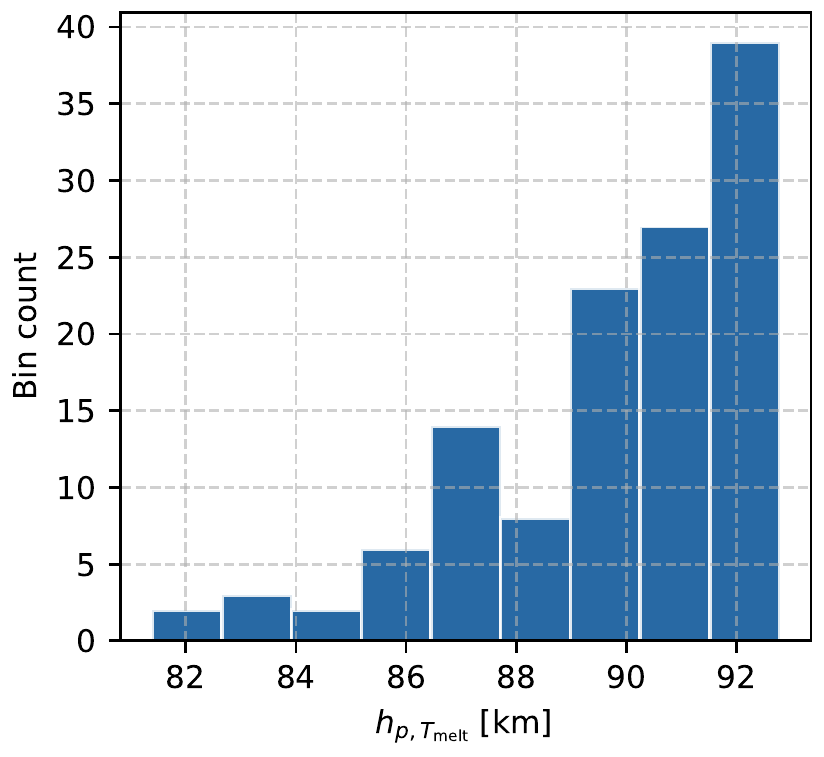}
		\caption{Distribution of the pericentre altitude at which the melting temperature, $T_{\rm melt}$, has been reached.}
		\label{fig:hist_tmelt}
	\end{subfigure}
	\hfill
	\begin{subfigure}[t]{0.45\textwidth}
		\centering
		\includegraphics[width=3.25in]{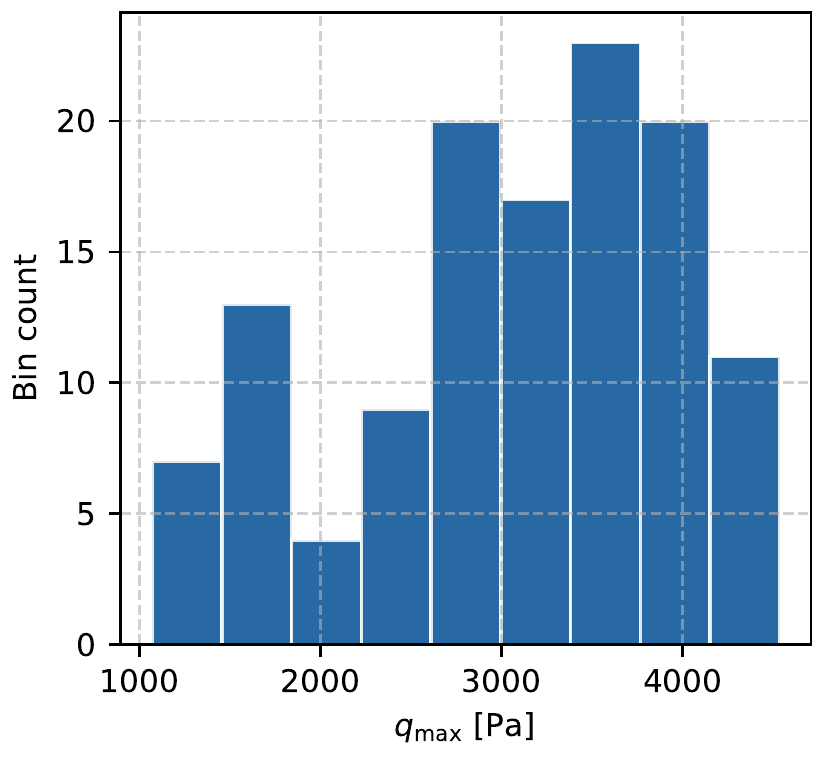}
		\caption{Distribution of the maximum dynamic pressure, $q_{\rm max}$, reached by the parent spacecraft.}
		\label{fig:hist_pdyn}
	\end{subfigure}
	\caption{Aerothermal loads on the parent spacecraft for selected GSO disposal trajectories.}
	\label{fig:hist_aerothermal_loads}
\end{figure}

\section{Conclusions and Discussion}
\label{sec:conclusions}
This paper shows a re-entry and demisability analysis of GSO satellites disposed on resonant trajectories. In fact, disposal through re-entry from GSO is possible for highly-inclined orbits exploiting lunisolar perturbations. These types of disposals are based on a mechanism of eccentricity build-up that eventually leads to re-entry. When propagating such trajectories with the PlanODyn long-term propagator, the contribution of drag lead to a substantial circularisation of the orbits before re-entry. This behaviour has been verified interfacing the long-term propagation with an object-oriented destructive re-entry tool (\emph{Phoenix}). In contrast to the long-term propagation, the \phoenix tool showed that the disposal trajectories lead to an eccentric re-entry: more than one-third of the re-entries occurs with an eccentricity higher than 0.8 and in 85\% of the cases the eccentricity is above 0.4. This is an interesting aspect as understanding the re-entry modes for these types of re-entry trajectories is important for a possible future exploitation. 
In an effort to find a more convenient way to interface the long-term propagation and the re-entry predictions, we borrowed the concept of overshoot boundary and applied it to a set of 130 disposal trajectories. The results showed that the overshoot boundary can be a viable option for the prediction of the entry conditions on GSO disposal trajectories. In fact, the re-entry predictions performed with the overshoot boundary closely matched the ones of the object-oriented code \emph{Phoenix}.
For a better understanding of the re-entry process from GSO resonant trajectories, we performed a demisability analysis with the \phoenix tool using a representative spacecraft configuration. The results showed that in no case the configuration was compliant with the current casualty risk regulations and that the application of design-for-demise techniques, or controlled and semi-controlled re-entries would be necessary to allow GSO satellites to be disposed via re-entry. Finally, we performed a preliminary analysis on the aerothermal loads suffered by the spacecraft while on the disposal trajectories and before the final re-entry phase. Because for most of the disposal trajectories the spacecraft performs several low-altitude passages before reaching the entry conditions, we observed that the spacecraft may experience high thermal loads. These loads may lead to an early breakup of the spacecraft and, therefore, must be further studied.

\section*{Acknowledgements}
This research has received funding from the European Research Council (ERC) under the European Union’s Horizon 2020 research and innovation programme (grant agreement No 679086 - COMPASS).

The authors would also like to thank Dr. Ioannis Gkolias for the help in the selection of the initial conditions of the long-term propagation of GSO orbits.

\bibliographystyle{model5-names}
\biboptions{authoryear}
\bibliography{mybib}

\begin{thebibliography}{40}
\expandafter\ifx\csname natexlab\endcsname\relax\def\natexlab#1{#1}\fi
\providecommand{\url}[1]{\texttt{#1}}
\providecommand{\href}[2]{#2}
\providecommand{\path}[1]{#1}
\providecommand{\DOIprefix}{doi:}
\providecommand{\ArXivprefix}{arXiv:}
\providecommand{\URLprefix}{URL: }
\providecommand{\Pubmedprefix}{pmid:}
\providecommand{\doi}[1]{\href{http://dx.doi.org/#1}{\path{#1}}}
\providecommand{\Pubmed}[1]{\href{pmid:#1}{\path{#1}}}
\providecommand{\bibinfo}[2]{#2}
\ifx\xfnm\relax \def\xfnm[#1]{\unskip,\space#1}\fi
\bibitem[{Alessi et~al.(2016)Alessi, Deleflie, Rosengren, Rossi, Valsecchi,
  Daquin \& Merz}]{alessi2016numerical}
\bibinfo{author}{Alessi, E.}, \bibinfo{author}{Deleflie, F.},
  \bibinfo{author}{Rosengren, A.}, \bibinfo{author}{Rossi, A.},
  \bibinfo{author}{Valsecchi, G.}, \bibinfo{author}{Daquin, J.}, \&
  \bibinfo{author}{Merz, K.} (\bibinfo{year}{2016}).
\newblock \bibinfo{title}{A numerical investigation on the eccentricity growth
  of gnss disposal orbits}.
\newblock {\it \bibinfo{journal}{Celestial Mechanics and Dynamical
  Astronomy}\/},  {\it \bibinfo{volume}{125}\/}\bibinfo{issue}{(1)},
  \bibinfo{pages}{71--90}. \DOIprefix\doi{10.1007/s10569-016-9673-4}.
\bibitem[{Armellin \& San-Juan(2018)}]{armellin2018optimal}
\bibinfo{author}{Armellin, R.}, \& \bibinfo{author}{San-Juan, J.~F.}
  (\bibinfo{year}{2018}).
\newblock \bibinfo{title}{Optimal earth’s reentry disposal of the galileo
  constellation}.
\newblock {\it \bibinfo{journal}{Advances in Space Research}\/},  {\it
  \bibinfo{volume}{61}\/}\bibinfo{issue}{(4)}, \bibinfo{pages}{1097--1120}.
  \DOIprefix\doi{10.1016/j.asr.2017.11.028}.
\bibitem[{Armellin et~al.(2015)Armellin, San-Juan \& Lara}]{armellin2015end}
\bibinfo{author}{Armellin, R.}, \bibinfo{author}{San-Juan, J.~F.}, \&
  \bibinfo{author}{Lara, M.} (\bibinfo{year}{2015}).
\newblock \bibinfo{title}{End-of-life disposal of high elliptical orbit
  missions: The case of integral}.
\newblock {\it \bibinfo{journal}{Advances in Space Research}\/},  {\it
  \bibinfo{volume}{56}\/}\bibinfo{issue}{(3)}, \bibinfo{pages}{479--493}.
  \DOIprefix\doi{10.1016/j.asr.2015.03.020}.
\bibitem[{Chapman(1959)}]{Chapman1959}
\bibinfo{author}{Chapman, D.~R.} (\bibinfo{year}{1959}).
\newblock {\it \bibinfo{title}{{An analysis of the corridor and guidance
  requirements for supercircular entry into planetary atmospheres}}\/}.
\newblock \bibinfo{type}{Technical Report} \bibinfo{number}{R-55} National
  Aeronautics and Space Administration.
\bibitem[{CIESIN(2020)}]{CIESIN2020}
\bibinfo{author}{CIESIN} (\bibinfo{year}{2020}).
\newblock \bibinfo{title}{{Gridded Population of the World v4 (GPWv4):
  population density}}.
\newblock \DOIprefix\doi{10.7927/H4NP22DQ}.
\bibitem[{Colombo(2019)}]{colombo2019long}
\bibinfo{author}{Colombo, C.} (\bibinfo{year}{2019}).
\newblock \bibinfo{title}{Long-term evolution of highly-elliptical orbits:
  luni-solar perturbation effects for stability and re-entry}.
\newblock {\it \bibinfo{journal}{Frontiers in Astronomy and Space Sciences}\/},
   {\it \bibinfo{volume}{6}\/}, \bibinfo{pages}{34}.
  \DOIprefix\doi{10.3389/fspas.2019.00034}.
\bibitem[{Colombo et~al.(2014{\natexlab{a}})Colombo, Letizia, Alessi \&
  Landgraf}]{Colombo2014integral}
\bibinfo{author}{Colombo, C.}, \bibinfo{author}{Letizia, F.},
  \bibinfo{author}{Alessi, E.~M.}, \& \bibinfo{author}{Landgraf, M.}
  (\bibinfo{year}{2014}{\natexlab{a}}).
\newblock \bibinfo{title}{{End-of-Life Earth Re-Entry for Highly Elliptical
  Orbits: the INTEGRAL Mission}}.
\newblock {\it \bibinfo{journal}{Advances in the Astronautical Sciences}\/},
  {\it \bibinfo{volume}{152}\/}, \bibinfo{pages}{1771--1791}.
\bibitem[{Colombo et~al.(2014{\natexlab{b}})Colombo, Letizia, Alessi, Landgraf
  et~al.}]{colombo2014end}
\bibinfo{author}{Colombo, C.}, \bibinfo{author}{Letizia, F.},
  \bibinfo{author}{Alessi, E.~M.}, \bibinfo{author}{Landgraf, M.} et~al.
  (\bibinfo{year}{2014}{\natexlab{b}}).
\newblock \bibinfo{title}{End-of-life earth re-entry for highly elliptical
  orbits: the integral mission}.
\newblock In {\it \bibinfo{booktitle}{Proceedings of the 24th AAS/AIAA Space
  Flight Mechanics Meeting}\/}.
\bibitem[{Colombo et~al.(2019)Colombo, Scala \& Gkolias}]{Colombo2019surf}
\bibinfo{author}{Colombo, C.}, \bibinfo{author}{Scala, F.}, \&
  \bibinfo{author}{Gkolias, I.} (\bibinfo{year}{2019}).
\newblock \bibinfo{title}{Surfing in the phase space of earths oblateness and
  third body perturbations}.
\newblock {\it \bibinfo{journal}{Advances in the Astronautical Sciences}\/},
  {\it \bibinfo{volume}{168}\/}\bibinfo{issue}{(AAS 19-484)},
  \bibinfo{pages}{3209--3227}.
\bibitem[{Colombo et~al.(2016)}]{colombo2016planetary}
\bibinfo{author}{Colombo, C.} et~al. (\bibinfo{year}{2016}).
\newblock \bibinfo{title}{Planetary orbital dynamics (planodyn) suite for long
  term propagation in perturbed environment}.
\newblock In {\it \bibinfo{booktitle}{6th International Conference on
  Astrodynamics Tools and Techniques (ICATT)}\/} (pp. \bibinfo{pages}{14--17}).
\bibitem[{Delhaise \& Morbidelli(1993)}]{delhaise1993luni}
\bibinfo{author}{Delhaise, F.}, \& \bibinfo{author}{Morbidelli, A.}
  (\bibinfo{year}{1993}).
\newblock \bibinfo{title}{Luni-solar effects of geosynchronous orbits at the
  critical inclination}.
\newblock {\it \bibinfo{journal}{Celestial Mechanics and Dynamical
  Astronomy}\/},  {\it \bibinfo{volume}{57}\/}\bibinfo{issue}{(1)},
  \bibinfo{pages}{155--173}. \DOIprefix\doi{10.1007/BF00692471}.
\bibitem[{{ESA Space Debris Mitigation
  WG}(2015)}]{ESASpaceDebrisMitigation2015}
\bibinfo{author}{{ESA Space Debris Mitigation WG}} (\bibinfo{year}{2015}).
\newblock {\it \bibinfo{title}{{ESA Space Debris Mitigation Compliance
  Verification Guidelines}}\/}.
\newblock \bibinfo{type}{Technical Report} \bibinfo{number}{ESSB-HB-U-002}
  European Space Agency.
\bibitem[{{European Space Agency}(2021)}]{DISCOSweb}
\bibinfo{author}{{European Space Agency}} (\bibinfo{year}{2021}).
\newblock \bibinfo{title}{{DISCOSweb}}.
\newblock \URLprefix \url{https://discosweb.esoc.esa.int/}
  \bibinfo{note}{(Accessed on 2021-01-04)}.
\bibitem[{Gkolias \& Colombo(2017)}]{gkolias2017end}
\bibinfo{author}{Gkolias, I.}, \& \bibinfo{author}{Colombo, C.}
  (\bibinfo{year}{2017}).
\newblock \bibinfo{title}{End-of-life disposal of geosynchronous satellites}.
\newblock In {\it \bibinfo{booktitle}{68th International Astronautical Congress
  (IAC 2017)}\/} (pp. \bibinfo{pages}{3613--3619}).
\newblock \bibinfo{organization}{International Astronautical Federation, IAF}
  volume~\bibinfo{volume}{6}.
\bibitem[{Gkolias \& Colombo(2019)}]{Gkolias2019GEO}
\bibinfo{author}{Gkolias, I.}, \& \bibinfo{author}{Colombo, C.}
  (\bibinfo{year}{2019}).
\newblock \bibinfo{title}{{Towards a sustainable exploitation of the
  geosynchronous orbital region}}.
\newblock {\it \bibinfo{journal}{Celestial Mechanics and Dynamical
  Astronomy}\/},  {\it \bibinfo{volume}{131}\/}\bibinfo{issue}{(4)},
  \bibinfo{pages}{1--30}. \DOIprefix\doi{10.1007/s10569-019-9895-3}.
  \href{http://arxiv.org/abs/1904.00473}{\tt arXiv:1904.00473}.
\bibitem[{Hicks(2009)}]{Hicks2009}
\bibinfo{author}{Hicks, K.~D.} (\bibinfo{year}{2009}).
\newblock {\it \bibinfo{title}{{Introduction to Astrodynamic Re-Entry}}\/}.
\newblock \bibinfo{type}{Technical Report} Air Force Institute of Technology.
\bibitem[{{IADC}(2011)}]{IADC2011}
\bibinfo{author}{{IADC}} (\bibinfo{year}{2011}).
\newblock \bibinfo{title}{{Inter-Agency Space debris Coordination Committee}}.
\newblock \URLprefix \url{https://www.iadc-home.org/#}.
\bibitem[{Jenkin \& McVey(2008)}]{jenkin2008lifetime}
\bibinfo{author}{Jenkin, A.}, \& \bibinfo{author}{McVey, J.}
  (\bibinfo{year}{2008}).
\newblock \bibinfo{title}{Lifetime reduction for highly inclined, highly
  eccentric disposal orbits by changing inclination}.
\newblock In {\it \bibinfo{booktitle}{AIAA/AAS Astrodynamics Specialist
  Conference and Exhibit}\/} (p. \bibinfo{pages}{7376}).
\newblock \DOIprefix\doi{10.2514/6.2008-7376}.
\bibitem[{Johnson \& Stansbery(2010)}]{johnson2010_nasa_debris}
\bibinfo{author}{Johnson, N.~L.}, \& \bibinfo{author}{Stansbery, E.~G.}
  (\bibinfo{year}{2010}).
\newblock \bibinfo{title}{{The new NASA orbital debris mitigation procedural
  requirements and standards}}.
\newblock {\it \bibinfo{journal}{Acta Astronautica}\/},  {\it
  \bibinfo{volume}{66}\/}\bibinfo{issue}{(3-4)}, \bibinfo{pages}{362--367}.
  \DOIprefix\doi{10.1016/j.actaastro.2009.07.009}.
\bibitem[{Klett(1964)}]{Klett1964}
\bibinfo{author}{Klett, R.~D.} (\bibinfo{year}{1964}).
\newblock {\it \bibinfo{title}{{Drag Coefficients and Heating Ratios for Right
  Circular Cylinder in Free-Molecular and Continuum Flow from Mach 10 to
  30}}\/}.
\newblock \bibinfo{type}{Technical Report} \bibinfo{number}{SC-RR-64-2141}.
\newblock \DOIprefix\doi{10.2172/4630398}.
\bibitem[{Kozai(1962)}]{kozai1962secular}
\bibinfo{author}{Kozai, Y.} (\bibinfo{year}{1962}).
\newblock \bibinfo{title}{Secular perturbations of asteroids with high
  inclination and eccentricity}.
\newblock {\it \bibinfo{journal}{The Astronomical Journal}\/},  {\it
  \bibinfo{volume}{67}\/}, \bibinfo{pages}{591--598}.
  \DOIprefix\doi{10.1086/108790}.
\bibitem[{Lidov(1963)}]{lidov1963evolution}
\bibinfo{author}{Lidov, M.} (\bibinfo{year}{1963}).
\newblock \bibinfo{title}{Evolution of the orbits of artificial satellites of
  planets as affected by gravitational perturbation from external bodies}.
\newblock {\it \bibinfo{journal}{AIAA Journal}\/},  {\it
  \bibinfo{volume}{1}\/}\bibinfo{issue}{(8)}, \bibinfo{pages}{1985--2002}.
  \DOIprefix\doi{10.2514/3.1983}.
\bibitem[{{MatWeb LLC}(2015)}]{MatWebLLC2015}
\bibinfo{author}{{MatWeb LLC}} (\bibinfo{year}{2015}).
\newblock \bibinfo{title}{{MatWeb - Material Property Data}}.
\newblock \URLprefix \url{http://www.matweb.com/}.
\bibitem[{Merz et~al.(2015)Merz, Krag, Lemmens, Funke, B{\"o}ttger, Sieg,
  Ziegler, Vasconcelos, Sousa, Volpp et~al.}]{merz2015orbit}
\bibinfo{author}{Merz, K.}, \bibinfo{author}{Krag, H.},
  \bibinfo{author}{Lemmens, S.}, \bibinfo{author}{Funke, Q.},
  \bibinfo{author}{B{\"o}ttger, S.}, \bibinfo{author}{Sieg, D.},
  \bibinfo{author}{Ziegler, G.}, \bibinfo{author}{Vasconcelos, A.},
  \bibinfo{author}{Sousa, B.}, \bibinfo{author}{Volpp, H.} et~al.
  (\bibinfo{year}{2015}).
\newblock \bibinfo{title}{Orbit aspects of end-of-life disposal from highly
  eccentric orbits}.
\newblock In {\it \bibinfo{booktitle}{Proceedings of the 25th International
  Symposium on Space Flight Dynamics ISSFD, Munich, Germany}\/}.
\bibitem[{{National Oceanic and Atmospheric
  Administration}(1976)}]{1976_std_atm}
\bibinfo{author}{{National Oceanic and Atmospheric Administration}}
  (\bibinfo{year}{1976}).
\newblock {\it \bibinfo{title}{{U.S. Standard Atmosphere 1976}}\/}.
\newblock \bibinfo{address}{Washington, D.C.}: \bibinfo{publisher}{National
  Oceanic and Atmospheric Administration, {\&} United States Air Force}.
\bibitem[{O'Connor(2008)}]{OConnor2008_NASA_handbook}
\bibinfo{author}{O'Connor, B.} (\bibinfo{year}{2008}).
\newblock {\it \bibinfo{title}{{Handbook for Limiting Orbital Debris}}\/}.
\newblock \bibinfo{type}{Technical Report} \bibinfo{number}{NASA-HDBK 8719.14}.
\bibitem[{Skoulidou et~al.(2017)Skoulidou, Rosengren, Tsiganis \&
  Voyatzis}]{skoulidou2017cartographic}
\bibinfo{author}{Skoulidou, D.~K.}, \bibinfo{author}{Rosengren, A.~J.},
  \bibinfo{author}{Tsiganis, K.}, \& \bibinfo{author}{Voyatzis, G.}
  (\bibinfo{year}{2017}).
\newblock \bibinfo{title}{Cartographic study of the meo phase space for passive
  debris removal}.
\newblock In {\it \bibinfo{booktitle}{Proceedings of the 7th European
  Conference on Space Debris, Darmstadt, Germany}\/}.
\bibitem[{Tauber \& Sutton(1991)}]{tauber1991stagnation}
\bibinfo{author}{Tauber, M.~E.}, \& \bibinfo{author}{Sutton, K.}
  (\bibinfo{year}{1991}).
\newblock \bibinfo{title}{Stagnation-point radiative heating relations for
  earth and mars entries}.
\newblock {\it \bibinfo{journal}{Journal of Spacecraft and Rockets}\/},  {\it
  \bibinfo{volume}{28}\/}\bibinfo{issue}{(1)}, \bibinfo{pages}{40--42}.
\bibitem[{Trisolini et~al.(2015)Trisolini, Lewis \& Colombo}]{TCH2015_IAC}
\bibinfo{author}{Trisolini, M.}, \bibinfo{author}{Lewis, H.~G.}, \&
  \bibinfo{author}{Colombo, C.} (\bibinfo{year}{2015}).
\newblock \bibinfo{title}{{Survivability and Demise Criteria for Sustainable
  Spacecraft Design}}.
\newblock In {\it \bibinfo{booktitle}{66th International Astronautical
  Conference}\/}.
\newblock \bibinfo{address}{Jerusalem}.
\bibitem[{Trisolini et~al.(2016)Trisolini, Lewis \& Colombo}]{THC2016_JSSE}
\bibinfo{author}{Trisolini, M.}, \bibinfo{author}{Lewis, H.~G.}, \&
  \bibinfo{author}{Colombo, C.} (\bibinfo{year}{2016}).
\newblock \bibinfo{title}{{Demise and Survivability Criteria for Spacecraft
  Design Optimization}}.
\newblock {\it \bibinfo{journal}{Journal of Space Safety Engineering}\/},  {\it
  \bibinfo{volume}{3}\/}\bibinfo{issue}{(2)}, \bibinfo{pages}{83--93}.
  \DOIprefix\doi{10.1016/S2468-8967(16)30023-4}.
\bibitem[{Trisolini et~al.(2017)Trisolini, Lewis \& Colombo}]{THC2017_IAC}
\bibinfo{author}{Trisolini, M.}, \bibinfo{author}{Lewis, H.~G.}, \&
  \bibinfo{author}{Colombo, C.} (\bibinfo{year}{2017}).
\newblock \bibinfo{title}{{Demisability and survivability multi-objective
  optimisation for preliminary spacecraft design}}.
\newblock In {\it \bibinfo{booktitle}{68th International Astronautical
  Congress}\/}.
\newblock \bibinfo{address}{Adelaide, Australia}.
\bibitem[{Trisolini et~al.(2018{\natexlab{a}})Trisolini, Lewis \&
  Colombo}]{TLC18Sensitivity}
\bibinfo{author}{Trisolini, M.}, \bibinfo{author}{Lewis, H.~G.}, \&
  \bibinfo{author}{Colombo, C.} (\bibinfo{year}{2018}{\natexlab{a}}).
\newblock \bibinfo{title}{{Demisability and survivability sensitivity to
  design-for-demise techniques}}.
\newblock {\it \bibinfo{journal}{Acta Astronautica}\/},  {\it
  \bibinfo{volume}{145}\/}, \bibinfo{pages}{357--384}.
  \DOIprefix\doi{10.1016/j.actaastro.2018.01.050}.
\bibitem[{Trisolini et~al.(2018{\natexlab{b}})Trisolini, Lewis \&
  Colombo}]{Trisolini2018_AESCTE}
\bibinfo{author}{Trisolini, M.}, \bibinfo{author}{Lewis, H.~G.}, \&
  \bibinfo{author}{Colombo, C.} (\bibinfo{year}{2018}{\natexlab{b}}).
\newblock \bibinfo{title}{{Spacecraft design optimisation for demise and
  survivability}}.
\newblock {\it \bibinfo{journal}{Aerospace Science and Technology}\/},  {\it
  \bibinfo{volume}{77}\/}, \bibinfo{pages}{638--657}.
  \DOIprefix\doi{10.1016/j.ast.2018.04.006}.
\bibitem[{Trisolini et~al.(2021)Trisolini, Lewis \&
  Colombo}]{Trisolini2021_JSSE}
\bibinfo{author}{Trisolini, M.}, \bibinfo{author}{Lewis, H.~G.}, \&
  \bibinfo{author}{Colombo, C.} (\bibinfo{year}{2021}).
\newblock \bibinfo{title}{{Constrained optimisation of preliminary spacecraft
  configurations under the design-for-demise paradigm}}.
\newblock {\it \bibinfo{journal}{Journal of Space Safety Engineering}\/},  {\it
  \bibinfo{volume}{8}\/}\bibinfo{issue}{(1)}, \bibinfo{pages}{63--74}.
  \DOIprefix\doi{10.1016/j.jsse.2021.01.005}.
\bibitem[{{Union of Concerned
  Scientists}(2017)}]{UnionofConcernedScientists2017}
\bibinfo{author}{{Union of Concerned Scientists}} (\bibinfo{year}{2017}).
\newblock \bibinfo{title}{{UCS Satellite Database}}.
\newblock \URLprefix \url{https://www.ucsusa.org/resources/satellite-database}
  \bibinfo{note}{(Accessed on 2021-01-04)}.
\bibitem[{{United Nations}(2019)}]{UnitedNations2019population}
\bibinfo{author}{{United Nations}} (\bibinfo{year}{2019}).
\newblock {\it \bibinfo{title}{{World Population Prospects 2019}}\/}.
\newblock \bibinfo{type}{Technical Report} United Nations \bibinfo{address}{New
  York}.
\bibitem[{Vallado(2013)}]{Vallado2013}
\bibinfo{author}{Vallado, D.~A.} (\bibinfo{year}{2013}).
\newblock {\it \bibinfo{title}{{Fundamentals of Astrodynamics and
  Applications}}\/}.
\newblock (\bibinfo{edition}{4th} ed.).
\newblock \bibinfo{publisher}{Microcosm}.
\bibitem[{Vinh et~al.(1980)Vinh, Busemann \& Culp}]{Vinh1980}
\bibinfo{author}{Vinh, N.~X.}, \bibinfo{author}{Busemann, A.}, \&
  \bibinfo{author}{Culp, R.~D.} (\bibinfo{year}{1980}).
\newblock {\it \bibinfo{title}{{Hypersonic and Planetary Entry Flight
  Mechanics}}\/}.
\newblock \bibinfo{address}{Ann Arbor, MI}: \bibinfo{publisher}{University of
  Michigan Press}.
\bibitem[{Wytrzyszczak et~al.(2007)Wytrzyszczak, Breiter \&
  Borczyk}]{wytrzyszczak2007regular}
\bibinfo{author}{Wytrzyszczak, I.}, \bibinfo{author}{Breiter, S.}, \&
  \bibinfo{author}{Borczyk, W.} (\bibinfo{year}{2007}).
\newblock \bibinfo{title}{Regular and chaotic motion of high altitude
  satellites}.
\newblock {\it \bibinfo{journal}{Advances in Space Research}\/},  {\it
  \bibinfo{volume}{40}\/}\bibinfo{issue}{(1)}, \bibinfo{pages}{134--142}.
  \DOIprefix\doi{10.1016/j.asr.2006.11.020}.
\bibitem[{Zhang et~al.(2017)Zhang, Zhao, Hou, Zhu, Wang, Sun \&
  Zhang}]{zhang2017long}
\bibinfo{author}{Zhang, M.-J.}, \bibinfo{author}{Zhao, C.-Y.},
  \bibinfo{author}{Hou, Y.-G.}, \bibinfo{author}{Zhu, T.-L.},
  \bibinfo{author}{Wang, H.-B.}, \bibinfo{author}{Sun, R.-Y.}, \&
  \bibinfo{author}{Zhang, W.} (\bibinfo{year}{2017}).
\newblock \bibinfo{title}{Long-term dynamical evolution of tundra-type orbits}.
\newblock {\it \bibinfo{journal}{Advances in Space Research}\/},  {\it
  \bibinfo{volume}{59}\/}\bibinfo{issue}{(2)}, \bibinfo{pages}{682--697}.
  \DOIprefix\doi{10.1016/j.asr.2016.10.016}.

\end{thebibliography}

\appendix

\section{GSO spacecraft configuration}
\label{sec:appendix_config}
This section contains the description of the spacecraft configuration of a sample GSO satellite. The main characteristics of the spacecraft are the ones summarised in \cref{tab:spacecraft}, while \cref{tab:spacecraft_config} shows the full configuration that has been provided as input to the \phoenix tool. The configuration has been derived from a preliminary design of a GSO satellite. \cref{tab:spacecraft_config} summarises the main properties of all the defined components, such as the geometrical shape, the material and the dimensions. In the \phoenix tool, both the thermal mass $m$ and the aerodynamic mass $m_{\rm aero}$ can be defined to distinguish between the mass that takes part in the ablation process (thermal mass) from the total mass of the component, which determines its aerodynamics (aerodynamic mass). The parameter $n_c$ specifies the number of instances of a component to be considered. Finally, the \emph{Parent} parameter is used to specify the relation between two components. The parent id identifies the component that contains the one specified; this is particularly useful when sub-components need to be defined. For example, a battery cell, \emph{Battery Cell 1}, with id 24 has as parent component the \emph{Battery Box 1} with id 13, which is specified in its \emph{Parent} field. Similarly, for all the primary components, the \emph{Parent} is the main spacecraft structure, with id 0. The configuration of \cref{tab:spacecraft_config} contains the main spacecraft components, with particular attention towards the critical ones in terms of demisability, such as reaction wheel assemblies, fuel and pressuriser tanks. With the definition of these components, we account for about 500 \si{\kilo\gram} of the spacecraft mass, out of the 1200 \si{\kilo\gram} of its total dry mass. Therefore, we could model about 42\% of the spacecraft mass. As it is difficult to model every part of a spacecraft configuration without knowing its detailed design, we decided to focus on the most critical and best-known components.

\clearpage
\onecolumn
\noindent
\begin{longtable}[c]{lccccccccccc}
	\caption{Spacecraft configuration used in the \phoenix tool destructive re-entry simulations. \label{tab:spacecraft_config}} \\
	\hline
	Part & ID & Parent & Shape & Material & $m$ & $l$ & $r$ / $w$ & $h$ & $t_s$ & $m_{\rm aero}$ & $n_c$ \\
	& & & & & \si{\kilo\gram} & \si{\meter} & \si{\meter} & \si{\meter} & \si{\milli\meter} & \si{\kilo\gram} & \\
	\hline
	GSO Sat 			& 0 	& - 	& Box 			& Al 6061-T6 	& - 	& 3 	& 1.7 	& 1.7 	& 3 	& 1200 	& -  \\
	Solar panel 		& 1 	& 0 	& Flat Plate 	& Al 6061-T6 	& 50 	& 10 	& 3 	& - 	& - 	& - 	& 1 \\
	RWA 1 				& 2 	& 0 	& Box 			& Al 6061-T6 	& - 	& 0.35 	& 0.35 	& 0.35 	& 2 	& - 	& 1 \\
	RWA 2 				& 3 	& 0 	& Box 			& Al 6061-T6 	& - 	& 0.35 	& 0.35 	& 0.35 	& 2 	& - 	& 1 \\
	RWA EBox			& 4 	& 0 	& Box 			& Al 6061-T6	& - 	& 0.1 	& 0.1 	& 0.1 	& 2 	& 5 	& 2 \\
	Thruster 			& 5 	& 0 	& Cylinder 		& Inconel-600 	& - 	& 0.12 	& 0.05 	& - 	& - 	& - 	& 12 \\
	Apogee Motor 		& 5 	& 0 	& Cylinder 		& Inconel-600 	& 1.8 	& 0.25 	& 0.1 	& - 	& - 	& - 	& 1 \\
	Thruster EBox 		& 6 	& 0 	& Box 			& Al 6061-T6	& - 	& 0.12 	& 0.12 	& 0.12 	& 2 	& 5 	& 2 \\
	Hyd Tank	 		& 7 	& 0 	& Cylinder 		& Ti 6Al4V 		& - 	& 0.9 	& 0.47 	& - 	& 1.6 	& - 	& 4 \\
	NTO Tank	 		& 8 	& 0 	& Cylinder 		& Ti 6Al4V 		& - 	& 0.546 & 0.173 & - 	& 1 	& - 	& 6 \\
	Pressurant Tank		& 9 	& 0 	& Sphere 		& SS AISI-316 	& - 	& - 	& 0.376 & - 	& 10 	& - 	& 1 \\
	Star Tracker		& 10 	& 0 	& Cylinder 		& Al 7075-T6  	& 0.47 	& 0.188	& 0.06 	& - 	& - 	& - 	& 2 \\
	IMU	 				& 11 	& 0 	& Box 			& Al 7075-T6	& 0.33 	& 0.088	& 0.063	& 0.047	& - 	& - 	& 2 \\
	Solar Array Mech. 	& 12 	& 0 	& Box 			& SS AISI-316	& 1.5 	& 0.122 & 0.11 	& 0.07 	& - 	& - 	& 2 \\
	Battery Box 1 		& 13 	& 0 	& Box 			& Al 6061-T6	& - 	& 0.45 	& 0.45 	& 0.25 	& 2 	& 5 	& 1 \\
	Battery Box 2 		& 14 	& 0 	& Box 			& Al 6061-T6	& - 	& 0.45 	& 0.45 	& 0.25 	& 2 	& 5 	& 1 \\
	OBC Platform 		& 15 	& 0 	& Box 			& Al 6061-T6	& - 	& 0.1 	& 0.1 	& 0.1 	& 2 	& 9 	& 2 \\
	OBC Payload 		& 16 	& 0 	& Box 			& Al 6061-T6	& - 	& 0.25 	& 0.2 	& 0.2 	& 2 	& 10 	& 2 \\
	TRX Box 			& 17 	& 0 	& Box 			& Al 6061-T6	& - 	& 0.3 	& 0.2 	& 0.2 	& 2 	& 9 	& 1 \\
	OBDH	 			& 18 	& 0 	& Box 			& Al 6061-T6	& - 	& 0.3 	& 0.3 	& 0.3 	& 2 	& 15 	& 2 \\
	Antenna Mech.	 	& 20 	& 0 	& Box 			& SS AISI-316	& 1.8 	& 0.15 	& 0.11 	& 0.07 	& - 	& - 	& 2 \\
	Antenna		 		& 21 	& 0 	& Cylinder 		& Al 7075-T6 	& 35 	& 0.01 	& 0.75 	& - 	& - 	& - 	& 2 \\
	RW 1				& 22	& 2		& Cylinder		& SS AISI-316	& 2.9 	& 0.09	& 0.1 	& -		& -		& -		& 4 \\
	RW 2				& 23	& 3		& Cylinder		& SS AISI-316	& 2.9 	& 0.09	& 0.1 	& -		& -		& -		& 4 \\
	Batt. Cell 1		& 24	& 13	& Box			& Al 6061-T6	& 0.135 & 0.07	& 0.06 	& 0.018	& -		& -		& 80 \\
	Batt. Cell 2		& 25	& 14	& Box			& Al 6061-T6	& 0.135 & 0.07	& 0.06 	& 0.018	& -		& -		& 80 \\
	TRX Coil			& 26	& 7		& Cylinder		& SS AISI-316	& 5		& 0.2	& 0.025 & -		& -		& -		& 1 \\
	\hline
\end{longtable}

\end{document}